\documentclass[journal]{IEEEtran}
\usepackage{cite}

\ifCLASSINFOpdf
\else
\fi
\usepackage{amsmath}

\usepackage{algorithm}
\usepackage{algorithmic}

\usepackage{amssymb}

\usepackage{amsfonts}
\usepackage{amsmath,amsfonts,amssymb,epsfig,subfigure,cite,url,multirow,booktabs}
\usepackage{afterpage,algorithm,algorithmic,mathrsfs}
\usepackage{bm}
\usepackage{color}

\newtheorem{lemma}{Lemma}

\newtheorem{remark}{Remark}

\newtheorem{corollary}{Corollary}
\usepackage{multirow}

\usepackage{stfloats}
\usepackage{dcolumn}
\newcolumntype{d}[1]{D{.}{.}{#1}}

\begin{document}
%
\title{Fixed Region Beamforming Using Frequency Diverse Subarray for Secure MmWave Wireless Communications }
%
%
%

\author{Yuanquan Hong,~\IEEEmembership{Student member,~IEEE,}
        Xiaojun Jing,
        Hui Gao,~\IEEEmembership{Senior Member,~IEEE,}
        Yuan He,~\IEEEmembership{Member,~IEEE}


\thanks{The authors are with the Key Laboratory of Trustworthy Distributed Computing and Service (BUPT), Ministry of Education, School of Information and Communication Engineering, Beijing University of Posts and Telecommunications, Beijing 100876, China (e-mail:hongyuanquan@bupt.edu.cn; huigao@bupt.edu.cn; jxiaojun@bupt.edu.cn; yuanhe@bupt.edu.cn).}%
}

\maketitle

\begin{abstract}
Millimeter-wave (mmWave) using conventional phased array (CPA) enables highly directional and fixed angular beamforming (FAB), therefore enhancing physical layer security (PLS) in the angular domain. However, as the eavesdropper is located in the direction pointed by the mainlobe of the information-carrying beam, information leakage is inevitable and FAB cannot guarantee PLS performance. To address this threat, we propose a novel fixed region beamforming (FRB) by employing a frequency diverse subarray (FDSA) architecture to enhance the PLS performance for mmWave directional communications. In particular, we carefully introduce multiple frequency offset increments (FOIs) across subarrays to achieve a sophisticated beampattern synthesis that ensures a confined information transmission only within the desired angle-range region (DARR) in close vicinity of the target user. More specifically, we formulate the secrecy rate maximization problem with FRB over possible subarray FOIs, and consider two cases of interests, i.e., without/with the location information of eavesdropping, both turn out to be NP-hard. For the unknown eavesdropping location case, we propose a seeker optimization algorithm to minimize the maximum sidelobe peak of the beampattern outside the DARR. As for the known eavesdropping location case, a block coordinate descend linear approximation algorithm is proposed to minimize the sidelobe level in the eavesdropping region. By using the proposed FRB, the mainlobes of all subarrays are constructively superimposed in the DARR while the sidelobes are destructively overlayed outside the DARR. Therefore, FRB takes prominent effect on confining information transmission within the DARR. Numerical simulations demonstrate that the proposed FDSA-based FRB can provide superior PLS performance over the CPA-based FAB.

\end{abstract}


%
\IEEEpeerreviewmaketitle

\section{Introduction}
%
%
%
%
\IEEEPARstart{S}{ecure} communication is one of the major requirements for wireless systems. Conventional security approach in wireless communications is encryption\cite{5751298,7355564} in the upper-layer with the assumption that the eavesdropper has limited computational capability. But the wide angular broadcasting nature of wireless medium makes wireless systems vulnerable to eavesdropping. Recently, physical layer security (PLS) is emerging as an alternative security paradigm that explores the randomness of the wireless channel to achieve confidentiality and authentication \cite{8006265,8335290,7762075}. PLS approaches have some unique advantages over the existing upper-layer methods. For example, no heavy upper-layer cryptographic algorithm is needed, no additional computing resource is required, and PLS is naturally immune to the brute force attack \cite{7467419,8283647,SecurityNature}. Aiming at wider bandwidth and hence higher rate of wireless data transmission, the move to higher frequency band, such as millimeter-wave (mmWave) band, has been actively standardized in the fifth generation (5G) cellular system \cite{6515173,6834753,4014597,7833078}. The beamforming technology based on multi-antenna phased array is regarded as a feasible solution to compensate the severe path-loss and absorption in mmWave communications\cite{6736750}. The feature of ultra-short wavelength of mmWave enables the very compact antenna array of a large number of elements, which further promotes the application of beamforming. Meanwhile, highly directional narrow beam is beneficial to secure mmWave data transmission, wherein the information leakage from the sidelobe is limited and the eavesdropping outside the mainlobe is very difficult. To this end, the eavesdropping is more likely to happen along the mainlobe direction in the 5G system with mmWave \cite{SecurityNature}.

\begin{figure}[!t]
\centering
\includegraphics[width=3.5in]{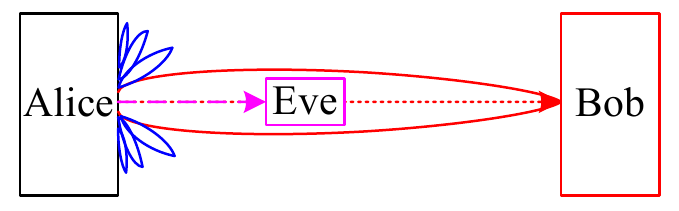}
\caption{PLS with conventional directional beamforming fails when Eve and Bob align with the mainlobe.}
\label{fig1}
\end{figure}

In recent years, the researches on PLS based narrow angular beamforming technologies have received more and more attention in mmWave wireless communications \cite{6051523,6781609,7081071,6941328,7880674,7964728,5159486,6544472,7386629,8334230}. The widely used conventional phased array (CPA) can enhance the transmission security with fixed angle beamforming (FAB), which reinforces the mainlobe of beam towards the desired receiver while suppressing the sidelobe in the undesired directions \cite{5159486,6544472,7386629}. In order to improve PLS performance, some FAB-based artificial noise (AN) methods based on CPA, such as antenna subset modulation (ASM)\cite{6544472}, switch phase array (SPA) transmission \cite{7386629,8334230,7417687}, inverted antenna subset transmission (IAST) \cite{8330319}, programmable weight phased-array transmission (PWPA) \cite{8329405}, are proposed to interfere the eavesdropper located in the sidelobe direction by scrambling the received constellation intentionally. These FAB methods achieve excellent PLS performance when the eavesdropper locates in the sidelobe direction. More recently, a frequency diverse array (FDA) \cite{8078202,8081593} is proposed to address the PLS problem when the legitimate user and eavesdropper are both located in the same direction pointed by the mainlobe with FAB. More specifically, an interesting rotated angular beamforming (RAB) scheme is proposed with FDA, where the resultant mainlobe can circumvent a well-localized eavesdropper via frequency-diverse beampattern synthesis. Although the RAB-resultant mainlobe can rotate and avoid one eavesdropper in a certain direction, it may still expose to the potential threat in another direction. Therefore, the success of PLS in existing CPA-based FAB and FDA-based RAB must rely on the assumption that the eavesdropper is not in the vicinity of the target user or far away from the mainlobe direction. As shown in Fig. \ref{fig1}, PLS performance will be degraded when the eavesdropper captures the mainlobe direction and moves in the beam direction.
\begin{table*}[!htbp]
\newcommand{\tabincell}[2]{\begin{tabular}{@{}#1@{}}#2\end{tabular}}
\caption{Comparison related work of PLS in phased array}
\centering
\begin{tabular}{|c|c|c|c|c|c|c|}
\hline Work& Array & Frequency &Sidelobe Constellation&Mainlobe Safety&Beam Parameters&PLS Countermeasure\\
\hline \tabincell{c}{\cite{5159486,6544472,7386629}}&\tabincell{c}{CPA}&\tabincell{c}{Single}&\tabincell{c}{Undistorted}&\tabincell{c}{$\times$} &\tabincell{c}{Angle}&\tabincell{c}{FAB}\\
\hline \tabincell{c}{\cite{6544472}\\ \cite{7386629}}&\tabincell{c}{ASM\\SPA}&\tabincell{c}{Single}&\tabincell{c}{Distorted}&\tabincell{c}{$\times$}&\tabincell{c}{Angle}&\tabincell{c}{FAB}\\
\hline \tabincell{c}{\cite{8330319,8329405}}&\tabincell{c}{PWPA}&\tabincell{c}{Single}&\tabincell{c} {Distorted}&\tabincell{c}{$\times$}&\tabincell{c}{Angle}&\tabincell{c}{FAB}\\
\hline \tabincell{c}{\cite{8078202,8081593}}&\tabincell{c}{FDA}&\tabincell{c}{Diverse}&\tabincell{c}{Undistorted}&\tabincell{c}{$\times$}&\tabincell{c}{Angle\ +\ Range}&\tabincell{c}{RAB}\\
\hline \tabincell{c}{Proposed}&\tabincell{c}{FDSA}&\tabincell{c}{Diverse}&\tabincell{c}{Distorted}&\tabincell{c}{$\surd$}&\tabincell{c}{Angle\ +\ Range}&\tabincell{c}{FRB}\\
\hline
\end{tabular}
\end{table*}

To the best of our knowledge, there is yet no effective method to address the eavesdropping threat along the mainlobe path. In this paper, we propose a novel fixed region beamforming (FRB) using frequency diverse subarray (FDSA) to address the mainlobe eavesdropping threat that is inherent in the FAB with CPA or RAB with FDA. The mainlobe generated by the proposed FRB focuses only in a desired angle-range region (DARR), where the subarray mainlobes are constructively superimposed within the DARR while the sidelobes are destructively overlayed outside the DARR. As a beneficial result, the signal-to-noise ratio (SNR) inside the DARR is much higher than the SNR outside the DARR, and the mainlobe path outside the DARR is eliminated. Therefore the proposed FRB can provide superior PLS performance over the FAB and RAB. More specifically, the proposed FDSA-based FRB exploits the degree-of-freedom (DoF) of frequency offset to generate an angle-range-dependent beampattern, which makes it possible to produce very low sidelobe in any region outside the DARR. In particular, we propose two algorithms to select the optimal frequency offset increment (FOI) vectors for secrecy rate maximization (SRM). Our proposed FRB distinguishes itself from the existing works from several important aspects as illustrated in Table \uppercase\expandafter{\romannumeral1}. Moreover, the key contributions and new features of this paper are summarized as follows.

\begin{itemize}
  \item \,\,We propose a novel FDSA-based FRB transmission scheme to enhance PLS for mmWave wireless communications. In contrast to the CPA-based FAB or FDA-based RAB that only generates an angle-confined beam, the proposed FRB can generate an angle-and-range double confined beam by exploiting the DoF of frequency offset with the FDSA. The sidelobe outside the DARR, even along the target direction, is attenuated dramatically by the destructive superposition. Therefore, the proposed FRB can address the mainlobe threat that is inherent in the CPA-based FAB and the FDA-based RAB.
  \item We propose a FRB optimization scheme without the location information of eavesdropping. In particular, we formulate the SRM problem which minimizes the maximum sidelobe peak outside the DARR. Aiming to solve the NP-hard problem, we propose a seeker optimization algorithm (SOA), which efficiently pursuits the optimized FOI vector across the subarray with fast convergence. The case study is of practical interests in the sense that the eavesdropper may hide itself and the worst-case optimization can guarantee basic PLS performance.
  \item We also propose a FRB optimization scheme with the location information of eavesdropping. In this case the objective of the SRM problem is to minimize the sidelobe level in the known eavesdropping location. Noting the problem is again NP-hard, we propose a block coordinate descend linear approximation algorithm (BCDLA) to efficiently find the optimized FOI vector. The case study is applicable to the practical scenario where the eavesdropper has been identified and localized.
  \item Comprehensive simulation results are offered to validate the advantages of the proposed FRB scheme. The secrecy rate (SR) is analyzed with different subarray parameters and maximum FOI. The problem of zero SR that is inherent in existing FAB and RAB is solved. The secrecy outage probability (SOP) is improved by 205 times with our proposed FDSA-based FRB over the existing CPA-based FAB when the SR threshold decrease to 1 bit/s/Hz.
\end{itemize}

\textbf{Organization}: The remainder of this paper is organized as follows. The system model is introduced in Section \uppercase\expandafter{\romannumeral2}. In Section \uppercase\expandafter{\romannumeral3}, we introduce the transmission technique of FRB. Subsequently, in Section \uppercase\expandafter{\romannumeral4} we analyze the algorithms of SR performance optimization. The simulation results are given Section \uppercase\expandafter{\romannumeral5}. Finally, we conclude our work in Section  \uppercase\expandafter{\romannumeral6}.

\textbf{Notation}: Boldface uppercase letters (e.g. \textbf{H}) represent matrix, boldface lowercase letters (e.g. \textbf{a}) denote vector, $a_i$ represents the $i$-th element of the vector \textbf{a}, $\mathcal{N}(\bm{\mu},\textbf{P})$ is a real Gaussian random vector with mean $\bm{\mu}$ and covariance \textbf{P}, $\mathcal{CN}(\bm{\mu},\textbf{P})$ is a complex Gaussian distribution with mean {$\bm{\mu}$ and covariance \textbf{P}. We use $[\cdot]^T$ and $[\cdot]^*$ to denote transpose, transpose and conjugate, respectively. We use $\mathbb{P}[\cdot]$, $\mathbb{E}[\cdot]$, $[\cdot]^+$ and $||\cdot||$  to denote probability, expectation, $\text{max}(0,[\cdot])$ and the Euclidean-norm, respectively. $\Re[\cdot]$ and $\Im[\cdot]$ denote real and imaginary, respectively. We use $\lfloor\cdot\rfloor$ to round the argument to the nearest integer towards $-\infty$.

\section{System Model}
In this section, the elements of the proposed FRB transmission will be introduced by presenting the transmitter architecture, system model and some directional beamforming schemes which are supportable by FRB.
\subsection{Transmitter Architecture}

As shown in Fig. \ref{fig2}, the transmitter consists of $M$ FDSAs plus frequency offset module at each branch. The bits of secret information are modulated in the baseband domain. Then the baseband signal is up-converted by $f_c$ through the mixer, and divided into $L$ signal copies. Each copy is converted again by an extra tiny frequency offset before phase shifting. Finally, the phase shifted signal is amplified by a power amplifier and loaded to the antenna.

For simplicity, we consider an uniform linear array (ULA) with an inter-element space of $d\leq 0.5\lambda$. The ULA is divided into $M$ non-overlapped FDSAs, denoted as ${S_0}$, $S_1$, $\ldots$, $S_{M-1}$, respectively. $\Delta f_{m,n}$ denotes the frequency offset in the $n$-th element of the $m$-th subarray. Each subarray consists of $N$ elements, i.e., the ULA has $L=MN$ elements. The frequency difference of adjacent elements in subarray $S_m$ is $\Delta f_m=\Delta f_{m,n+1}-\Delta f_{m,n}$, which is defined as FOI in subsequent section. Note the equal difference feature of frequency offset is helpful to reduce the complexity of beampattern optimization algorithm.

\subsection{System Model }
We consider a multiple-input single-output (MISO) mmWave wireless communication system with a $L$-antenna transmitter and a single antenna receiver in the line-of-sight (LoS) environment. Both target user (Bob) and eavesdropper (Eve) are each equipped with one antenna. The transmitter (Alice) knows the exact location $(\theta_T, R_T)$ of Bob but may not know the location of Eve in general. For simplicity, we only consider the directional LoS channel since the multi-path components (MPCs) are very few and relatively weak in mmWave transmission \cite{Smulders1994Deterministic,7876781,6544472,7386629}, which can be ignored. This assumption is acceptable in LoS mmWave communications because the MPCs are often attenuated by 20dB as compared to the LoS component \cite{8078202,8081593,mmWaveMPCs}.

As we assume the ULA is positioned on the x-y plane, the receiver direction can be identified by the azimuth angle $\theta$ only. The received signal at any location $(\theta, R)$ is written as
\begin{align}
y(\theta,R)&=\sqrt{P_T}\mathbf{h}^*(\theta,R)\mathbf{w}(\theta_T,R_T)x+\eta\nonumber\\
&=\sqrt{P_T}B(\theta,R)x+\eta\label{EquModel},
\end{align}
where $P_T$ is the transmission power, $\mathbf{h}\in\mathbb{C}^{L\times 1}$ is channel vector, $\textbf{w}\in\mathbb{C}^{L\times 1}$ is the transmit beamforming vector, $B=\mathbf{h}^*\mathbf{w}$ denotes the array factor at location $(\theta, R)$,  $x$ is the modulated transmit signal with $\mathbb{E}[|x|^2]=1$, and $\eta$ is the additive white gaussian noise (AWGN) with zero mean and $\sigma^2$ variance, i.e., $\eta\sim \mathcal{N}(0,\sigma^2)$.
\begin{figure}[!t]
\centering
\includegraphics[width=3.5in]{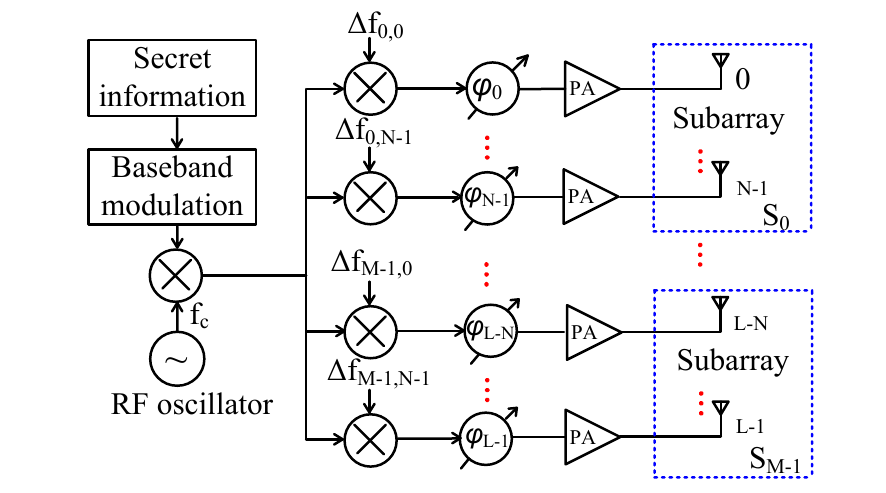}
\caption{The proposed FDSA-based FRB transmitter. It consists of $M$ FDSAs with a FOI vector $\Delta\mathbf{f}=[\Delta f_0,\cdots,\Delta f_{M-1}]^T\in\mathbb{R}^{M\times1}$. Each subarray has $N$ elements.}
\label{fig2}
\end{figure}
\subsection{Directional Beamforming Techniques}
Based on the proposed FDSA architecture, we can perform two different directional beamforming techniques by adjusting the FOIs, which are discussed as follows.

\emph{1) Fixed Angular Beamforming (FAB)}

The FDSA architecture in Fig. \ref{fig2} is simplified to a CPA architecture when the extra tiny frequency offset is zero, i.e. $\Delta f_{m,n}=0$, $\forall m,n$. In this special case, the simplified FDSA can also support FAB along the target direction with $L$ elements. The channel vector in this case can be written as \cite{Trees2002Detection}
\begin{align}
\mathbf{h}^*(\theta,R)&=\sqrt{P_L(R)}\mathbf{a}^{*}(\theta)\nonumber\\
&=\sqrt{P_L(R)}[e^{-j(\frac{L-1}{2})\frac{2\pi d}{\lambda}\cos\theta},e^{-j(\frac{L-1}{2}-1)\frac{2\pi d}{\lambda}\cos\theta},\nonumber\\
&\ \ \ \cdots,e^{j(\frac{L-1}{2})\frac{2\pi d}{\lambda}\cos\theta}],
\end{align}
where $\sqrt{P_L(R)}$ is the path-loss factor at the range of $R$, $\mathbf{a}(\theta)$ is array steering vector, and $\lambda$ is wavelength.

In order to highlight the beamforming gain, we temporarily ignore the path-loss factor for clarity. The transmitter can steer its mainlobe along $\theta_T$ by setting $\mathbf{w}=\frac{\mathbf{a}(\theta_T)}{L}$. Then the array factor of FAB along any direction $\theta$ can be written as
\begin{align}
B_{FAB}(\theta)&=\frac{1}{L}\sum\limits_{l=0}^{L-1}e^{-j(\frac{L-1}{2}-l)\frac{2\pi d}{\lambda}(\cos\theta-\cos\theta_T)}\nonumber\\
&=\frac{\sin(\frac{\pi Ld(\cos\theta-\cos\theta_T)}{\lambda})}{L\sin(\frac{\pi d(\cos\theta-\cos\theta_T)}{\lambda})}\nonumber\\
&=
\begin{cases}
1,\mathrm{if}\ \theta=\theta_T\\
<1,\mathrm{otherwise}
\end{cases}.
\end{align}

Since the array factor is constant along the direction $\theta$, the beampattern of FAB is range-independent. As shown in Fig. \ref{fig4}, the beampattern is highly directional. However, there is an inherent mainlobe threat for FAB. More specifically, conventional directional beamforming based PLS is effective when Eve is located outside the mainlobe angular sector, but it will be entirely ineffective when Eve locates inside the mainlobe angular sector, i.e., $\theta\approx\theta_T$, even Eve is far away from Bob. Intuitively, if the mainlobe pattern can be controlled to \emph{circumvent} the eavesdropper, the security threat in the target direction may be solved.
\begin{figure}[!t]
\centering
\includegraphics[width=3.5in]{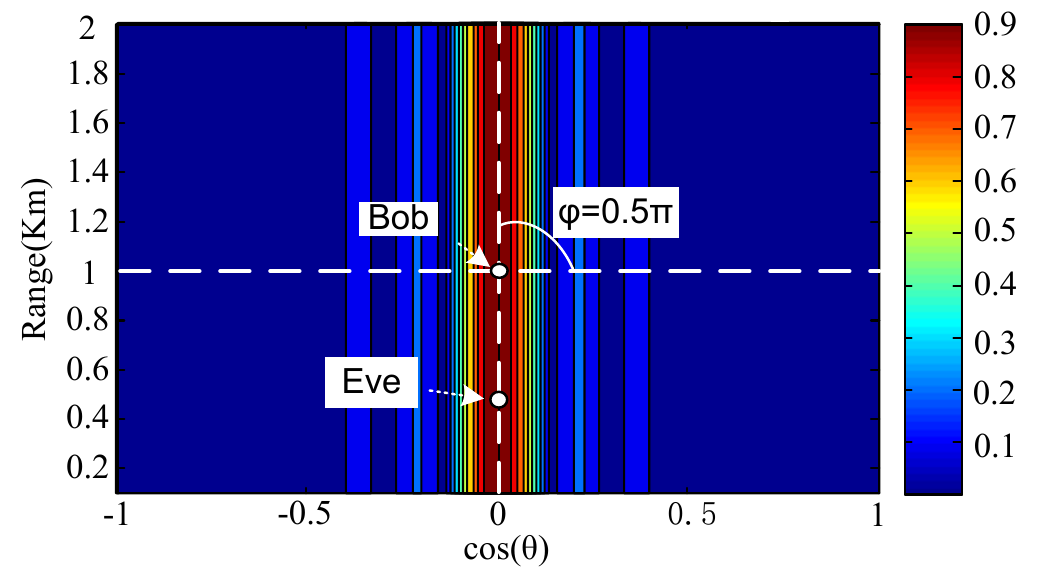}
\caption{Typical beampattern of CPA-based FAB with $f_c=60$GHz, $L=15$, $d=0.5\lambda$, $R_T=1$Km, $\theta_T=90^0$.}
\label{fig4}
\end{figure}

\emph{2) Rotated Angular Beamforming (RAB)}

In Fig. \ref{fig2}, assume that the frequency difference between adjacent elements is constant, i.e., $\Delta f=\Delta f_{m,n+1}-\Delta f_{m,n}=\Delta f_{m+1,0}-\Delta f_{m,N-1}$, $\forall m,n$, the FDSA will become an uniform linear FDA with $L$ elements and one FOI $\Delta f$. Thus, the radiated frequency of the $l$-th antenna in the FDA is expressed as
\begin{equation}
f_l=f_c-\Big(\frac{L-1}{2}-l\Big)\Delta f,l=0,1,\ldots,L-1,
\end{equation}
where $f_c$ denotes the carrier frequency. Thus, the frequency vector of FDA is expressed as
\begin{align}
\mathbf{f}=
\begin{cases}
[\cdots,f_c-\Delta f,f_c,f_c+\Delta f,\cdots]^T,\mathrm{odd}\ L\\
[\cdots,f_c-\frac{\Delta f}{2},f_c+\frac{\Delta f}{2},\cdots]^T,\mathrm{even}\ L
\end{cases}.
\end{align}

 Recall $d$ as the uniform antenna spacing of the FDA, which is set as $d\leq c/[2(f_c+(N-1)\triangle f)]\approx c/(2f_c)=0.5\lambda$ to avoid aliasing effects with $c$ being the speed of light.

Without loss of generality, we choose the array center, i.e., the origin of the coordinates, as the phase reference point. The location $(\theta, R)$ is a far-field point far away from the array center. Thus, the propagation range from the $l$-th element to the location $(\theta, R)$ is expressed as
\begin{equation}
R_l=R+\Big(\frac{L-1}{2}-l\Big)d\cos\theta.
\end{equation}

The phase difference between the $l$-th element and the reference is written as
\begin{align}\label{pdfda}
\Delta\Phi_l&=\frac{2\pi f_lR_l}{c}-\frac{2\pi f_cR}{c}\nonumber\\
&=\frac{2\pi (\frac{L-1}{2}-l)f_cd\cos\theta}{c}-\frac{2\pi(\frac{L-1}{2}-l)R\Delta f}{c}\nonumber\\
&\quad -\frac{2\pi(\frac{L-1}{2}-l)^2\Delta f d\cos\theta}{c}.
\end{align}
Note that in (\ref{pdfda}), the expression is identical to CPA-based FAB when $\Delta f=0$. The second term on the righthand side (RHS) of (\ref{pdfda}) shows that the phase difference of RAB is range-dependent and the third term can be neglected when $0.5(L-1)\Delta f\ll f_c$ \cite{7740083,6404099,6786333,Xu2017Range}. The channel vector of RAB can be written as
\begin{align}\label{EquASFDA}
&\mathbf{h}_{RAB}^*(\theta,R)=\sqrt{P_L(R)}\mathbf{a}^*(\theta,R)
\end{align}
where $\mathbf{a}(\theta,R)$ is array steering vector of RAB, which is defined as
\begin{align}
\mathbf{a}^{*}(\theta,R)=&[e^{-j\frac{N-1}{2}\frac{2\pi}{c}(f_cd\cos\theta-R\Delta f)},e^{-j\frac{N-3}{2}\frac{2\pi}{c}(f_cd\cos\theta-R\Delta f)}\nonumber\\
&,\cdots,e^{j\frac{N-1}{2}\frac{2\pi}{c}(f_cd\cos\theta-R\Delta f)}].
\end{align}

\begin{figure}[!t]
\centering
\includegraphics[width=3.5in]{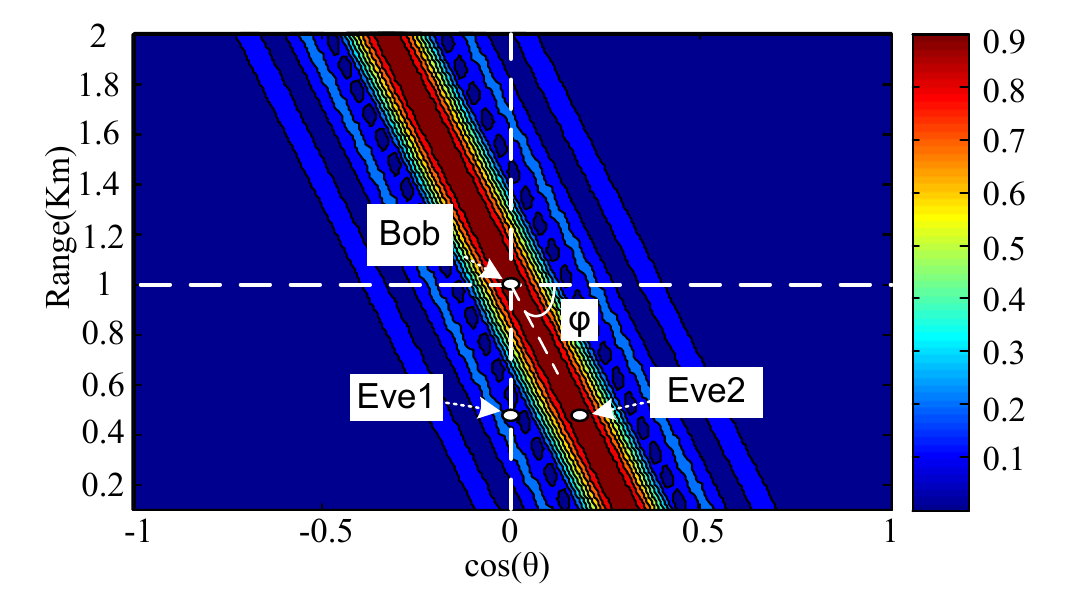}
\caption{Typical beampattern of FDA-based RAB with $f_c=60$GHz, $L=15$, $d=0.5\lambda$, $R_T=1$Km, $\theta_T=90^0$, $\Delta f=-45$KHz.}
\label{fig5}
\end{figure}

When we ignore path-loss factor for simplicity, the array factor of FDA-based RAB is calculated as
\begin{align}\label{ARFDA}
B_{RAB}(\theta,R)&=\mathbf{h}_{RAB}^*(\theta,R)\mathbf{w}_{RAB}(\theta_T,R_T)\nonumber\\
&=\frac{\sin\left(\frac{L\pi f_cd(\cos\theta-\cos\theta_T)}{c}-\frac{L\pi\Delta f(R-R_T)}{c}\right)}{L\sin\left(\frac{\pi f_cd(\cos\theta-\cos\theta_T)}{c}-\frac{\pi\Delta f(R-R_T)}{c}\right)}\nonumber\\
&=\frac{\sin(La-Lb\Delta f)}{L\sin(a-b\Delta f)}\nonumber\\
&=
\begin{cases}
1,\mathrm{if}\ \theta=\theta_T,R=R_T\\
<1,\mathrm{otherwise}
\end{cases},
\end{align}
where $a=\frac{\pi f_cd(\cos\theta-\cos\theta_T)}{c}$, $b=\frac{\pi (R-R_T)}{c}$, $\mathbf{w}_{RAB}(\theta_T,R_T)=\frac{\mathbf{a}(\theta_T,R_T)}{L}$ is the steering weight vector, and $(\theta_T,R_T)$ denotes the target point, i.e., Bob's location. Notice that FDA-based RAB can be simplified to CPA-based FAB by setting $\Delta f=0$. According to (\ref{ARFDA}), we know that the beampattern of RAB is angle-range-dependent while FAB is angle-dependent and range-independent.

As shown in Fig. \ref{fig5}, the mainlobe pattern of RAB can circumvent the eavesdropper by controlling the rotated angle $\varphi$ with FOI $\Delta f$. The rotating law of RAB is summarized in the following lemma.
\begin{lemma}
Assume that $\Delta f\ll f_c$  in the transmit beampattern of RAB in mmWave wireless communications, the pattern will be rotated around the target point $(\theta_T,R_T)$ in the $\cos\theta-R$ plane. The rotated angle $\varphi$ between the mainlobe center and $\cos\theta$-axis can be expressed as
\begin{equation}
\varphi=\arctan\frac{f_cd}{\Delta f}.
\end{equation}

\emph{Proof:} See Appendix A.$\hfill\blacksquare$
\end{lemma}

According to Lemma 1, we have some interesting observations which are summarized by the following corollaries.
\begin{corollary}
If $\Delta f=0$, the transmit beampattern of RAB only depends on the angle parameter $\theta$ and the range parameter $R$ is invalid.

\emph{Proof:} See Appendix B.$\hfill\blacksquare$
\end{corollary}
\begin{corollary}
If $\Delta f_1$ and $\Delta f_2$ are the FOIs of FDA, and $\Delta f_1=-\Delta f_2\neq 0$, then, the $\Delta f_1$-resultant beampattern is symmetric with the $\Delta f_2$-resultant beampattern along the $R$-axis.

\emph{Proof:} See Appendix C. $\hfill\blacksquare$
\end{corollary}
\begin{remark}
According to Lemma 1, the narrow angular mainlobe path of RAB can be rotated flexibly around the target point by controlling the FOI $\Delta f$, which makes it possible to circumvent the eavesdropper with known location. As shown in Fig. \ref{fig5}, the mainlobe path circumvents Eve1 who is located in the target direction $\theta_T$, and thus the PLS threat in the target direction is eliminated.
\end{remark}

However, RAB is still confronted by another security threat, because the mainlobe is only rotated to another direction which might point to another eavesdropper, such as Eve2 in Fig. \ref{fig5}. Noting the vulnerability of FAB, we propose FRB, a new beamforming solution, which is able to confine the mainlobe within a narrow angle-range region centered by the target user, namely the DARR, and suppresses the mainlobe gain outside the DARR for better PLS performance. In the next section, we will give detailed descriptions for FRB.

\section{FRB Transmission Technique}
In this section, we will introduce the beampattern and new features of the FDSA-based FRB. The feature of mainlobe suppressing outside DARR is characterized and the feature of maximum sidelobe peak is also summarized according the subbeam distribution, both serve as the preparations for PLS performance enhancement with FRB.
\subsection{Frequency Diverse Subarray}
\begin{figure}[!t]
\centering
\includegraphics[width=3.5in]{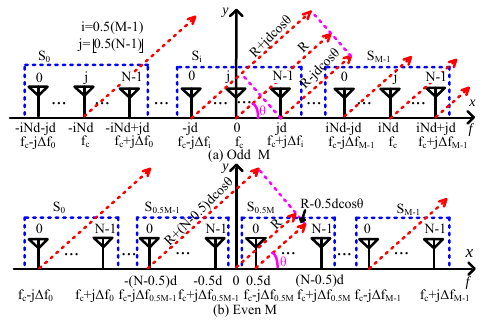}
\caption{Typical linear antenna array with $M$ FDSAs, uniformly distributed along the $x$-axis with $d\leq0.5\lambda$ spacing and centered at the origin. Each subarray consists of $N$ elements. (a) $M$ is odd, (b) $M$ is even.}
\label{fig3}
\end{figure}

As shown in Fig. \ref{fig3}, we divide the $L$ elements of ULA into $M$ non-overlapped FDSAs denoted as $S_{0}$, $S_{1}$,$\cdots$,$S_{M-1}$, respectively. Each FDSA has $N$ elements. The corresponding FOIs to these subarrays are $\Delta f_0$, $\Delta f_1$, $\cdots$, $\Delta f_{M-1}$, respectively.

The frequency of the $n$-th element in subarray $S_m$ is expressed as
\begin{align}
f_{m,n}=f_c-\Big(\frac{N-1}{2}-n\Big)\Delta f_m.
\end{align}
The propagation range from the $n$-th element of subarray $S_m$ to a far-field receiver is written as
\begin{align}
R_{m,n}=R+\Big[\Big(\frac{M-1}{2}-m\Big)N+\frac{N-1}{2}-n\Big]d\cos\theta.
\end{align}
Thus, the phase difference $\Delta \Phi_{m,n}$ between the $n$-th element of subarray $S_m$ and the array center is expressed as
\begin{align}\label{pdfab}
\Delta \Phi_{m,n}&=\frac{2\pi f_{m,n}R_{m,n}}{c}-\frac{2\pi f_cR}{c}\nonumber\\
&=\frac{2\pi(\frac{N-1}{2}-n)f_cd\cos\theta}{c}-\frac{2\pi (\frac{N-1}{2}-n)R\Delta f_m}{c}+\nonumber\\
&\quad\frac{2\pi(\frac{M-1}{2}-m)f_cNd\cos\theta}{c}-\frac{2\pi\Delta f_md\cos\theta}{c}\times\nonumber\\
&\quad\Big(\frac{N-1}{2}-n\Big)\Big[\Big(\frac{M-1}{2}-m\Big)N-\frac{N-1}{2}-n\Big].
\end{align}
Note that the first two terms on the RHS of (\ref{pdfab}) are equal to phase difference of FDA-based RAB, which are also angle-range-dependent. The third term denotes the constant part of phase difference in the $m$-th subarray $S_m$, which plays an important role in the subbeam superposition of FRB. The last term has a vary small value given the assumption $\Delta f_m\ll f_c$, and is far less important than the first term. Therefore, we neglect the last term in the subsequent section\cite{6404099,6786333,Xu2017Range}.
\subsection{FRB-Resultant Beampattern}
The channel vector of the proposed FRB is expressed as
\begin{align}\label{AFFAB}
&\mathbf{h}_{FRB}(\theta,R)=\sqrt{P_L(R)}\mathbf{a}_{FRB}(\theta,R)\nonumber\\
&=\sqrt{P_L(R)}
\left[
\begin{array}{c}
e^{j\frac{2\pi(\frac{M-1}{2})f_cNd\cos\theta}{c}}\mathbf{a}_{0}(\theta,R)\\
e^{j\frac{2\pi(\frac{M-1}{2}-1)f_cNd\cos\theta}{c}}\mathbf{a}_{1}(\theta,R)\\
\vdots\\
e^{-j\frac{2\pi(\frac{M-1}{2})f_cNd\cos\theta}{c}}\mathbf{a}_{M-1}(\theta,R)\\
\end{array}
\right],
\end{align}
where $\mathbf{a}_{FRB}(\theta,R)$ is composite steering vector of FRB, and $\mathbf{a}_m(\theta,R)$ represents the corresponding steering vector of the $m$-th subarray with FOI $\Delta f_m$, which is formulated by (\ref{EquASFDA}). When we ignore the path-loss factor, the channel vector is $\mathbf{h}_{FRB}(\theta,R)=\mathbf{a}_{FRB}(\theta,R)$.
\begin{figure}[!t]
\centering
\includegraphics[width=3.5in]{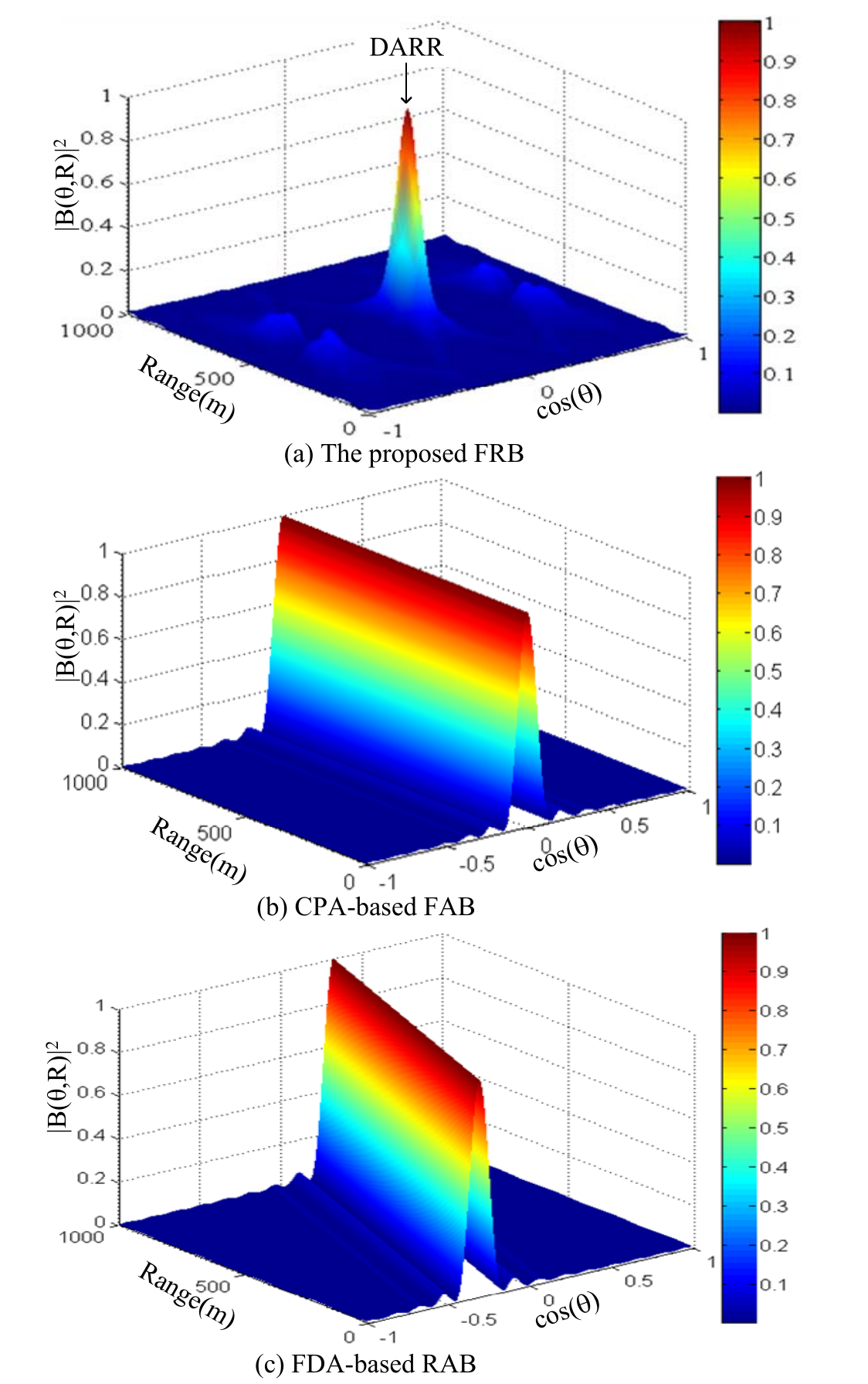}
\caption{Typical beampattern of three different techniques with $f_c=60$GHz, $\theta_T=90^0$, $R_T=500$m. (a) The proposed FRB with $N=15$, $M=10$, $\Delta \mathbf{f}=[-250, -200\cdots, -50, 50, 100, \cdots, 250]$KHz. (b) FAB with $L=15$. (c) RAB with $L=15$, $\Delta f=50$KHz.}
\label{fig6}
\end{figure}

Assuming that the location of the target user is $(\theta_T, R_T)$, and the corresponding transmit weight vector is  $\mathbf{w}_{FRB}(\theta_T,R_T)=\mathbf{a}_{FRB}(\theta_T,R_T)/MN$. Thus, the FRB-resultant beampattern can be expressed as
\begin{align}\label{bfam}
&B_{FRB}(\theta,R)=\mathbf{h}_{FRB}^*(\theta,R)\mathbf{w}_{FRB}(\theta_T,R_T)\nonumber\\
&=\sum\limits_{m=0}^{M-1}\frac{e^{j\frac{2\pi(\frac{M-1}{2}-m)f_cNd(\cos\theta-\cos\theta_T)}{c}}\mathbf{a}_m^*(\theta,R)\mathbf{a}_m(\theta_T,R_T)}{MN}\nonumber\\
&=\sum\limits_{m=0}^{M-1}\frac{e^{j(M-1-2m)Na}\sin(Na-Nb\Delta f_m)}{MN\sin(a-b\Delta f_m)}\nonumber\\
&=\sum\limits_{m=0}^{M-1}\frac{\rho(\theta)\sin(Na-Nb\Delta f_m)}{MN\sin(a-b\Delta f_m)},
\end{align}
where $a=\frac{\pi f_cd(\cos\theta-\cos\theta_T)}{c}$, $b=\frac{\pi (R-R_T)}{c}$. Here $\rho(\theta)=e^{j(M-1-2m)Na}$ is the phase shift factor of the resultant beampattern which has constructive or destructive effect on the superposition of $M$ subbeams.

The amplitude of the FRB-resultant beampattern can be written as
\begin{align}
|B_{FRB}(\theta,R)|=
\begin{cases}
1 ,\text{if $R=R_T$, $\theta=\theta_T$ } \\
<1,\text{otherwise}.
\end{cases}
\end{align}

Fig. \ref{fig6}(a) demonstrates the amplitude distribution of the proposed FRB in angle-range domain. Inside the DARR, the amplitude of the FRB-resultant beampattern reaches the maximum peak, which can provide very high received array gain for Bob. On the other hand, the amplitude is attenuated remarkably outside the DARR. In contrast to the pulse-shape beampattern of FRB in Fig. \ref{fig6}(a), FAB in Fig. \ref{fig6}(b) and RAB in Fig. \ref{fig6}(c) all show the belt-shape beampattern in the angle-range domain, which implies the conceivable mainlobe threat.

\subsection{The amplitude feature of FRB-resultant beampattern}
Define $\Delta \mathbf{f}=[\Delta f_0,\cdots,\Delta f_{M-1}]$ as the FOI vector of the FDSA-based FRB. According to Lemma 1, we can rotate the subbeam's mainlobe path towards the desired direction by controlling the FOI. As shown in Fig. \ref{fig7}, the blue lines are the mainlobe paths of FDSAs. The whole plane can be partitioned into three regions according to the distribution of subarray mainlobe \cite{Xu2017Range}, denoted with $\Omega_m$, $\Omega_{ms}$ and $\Omega_s$, respectively. In Fig. \ref{fig7}, we use green, blue and red to mark $\Omega_m$, $\Omega_{ms}$ and $\Omega_s$, respectively. More specifically, in $\Omega_m$, all the subbeam mainlobes of the FDSA are traversed, and the phase shift factors are all equal to 1 which have constructive effect on the superposition of those subbeam's mainlobes. Thus, $\Omega_m$ represents the mainlobe region of the FRB-resultant beampattern, i.e., the DARR.

The remaining region is the sidelobe of the FRB-resultant beampattern. $\Omega_{ms}$ consists of $K$ mainlobe(s)\footnote{$1\leq K \leq M-1$ and $K\in\mathbb{N}$. But there exists only one mainlobe, i.e., $K=1$, for the most part in $\Omega_{ms}$.} and $(M-K)$ sidelobe(s) at the same time, and in $\Omega_s$ $M$ sidelobes exist. The phase shift factor $\rho(\theta)$ in (\ref{bfam}) is a complex number which has destructive contribution to the superposition of subbeams and generates very low gain in this region.

\begin{figure}[!t]
\centering
\includegraphics[width=3.5in]{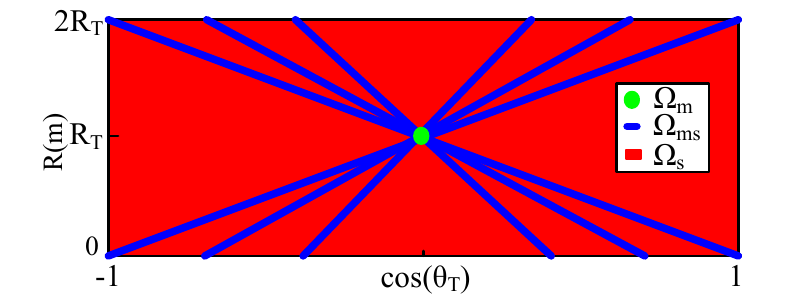}
\caption{The partition of resultant beampattern in the proposed FRB with $M=6$. The blue lines are the mainlobe paths of FDSA. The region marked by green, blue and red denotes $\Omega_m$, $\Omega_{ms}$ and $\Omega_s$, respectively }
\label{fig7}
\end{figure}

\begin{lemma}
 Let $B_{\Omega_m}^p$, $B_{\Omega_{ms}}^p$ and $B_{\Omega_s}^p$ denote the maximum peak levels in the regions $\Omega_m$, $\Omega_{ms}$ and $\Omega_s$, respectively. Then, we have
\begin{equation}\label{lem2}
B_{\Omega_m}^p\geq B_{\Omega_{ms}}^p\geq B_{\Omega_s}^p.
\end{equation}

\emph{Proof:} See Appendix D.$\hfill\blacksquare$
\end{lemma}

\begin{corollary}\label{coro3}
The maximum sidelobe peak ($\cos\theta_n,R_m$) of the FRB-resultant beampattern is confined in the $\Omega_{ms}$ region with
\begin{align}
\cos\theta_n=
\begin{cases}
\frac{\lambda n}{Nd},n=0,\pm1,\cdots,\lfloor\frac{Nd}{\lambda}\rfloor,\mathrm{Odd}\ M \\
\frac{\lambda n}{2Nd},n=0,\pm1,\cdots,\lfloor\frac{2Nd}{\lambda}\rfloor,\mathrm{Even}\ M
\end{cases},
\end{align}
\begin{align}
R_m=\frac{f_cd}{\Delta f_m}\cos\theta_n,m=0,1,\cdots,M-1.\ \ \  \ \ \ \ \ \ \ \
\end{align}

\emph{Proof:} See Appendix E.$\hfill\blacksquare$
\end{corollary}

From (\ref{bfam}), we know that the sidelobe peak of FRB is controlled by three parameters, namely, angle, range and FOIs. Furthermore, the location of sidelobe peak varies with the FOIs, which brings extra difficulties in the beampattern optimization in next section.

Maximum sidelobe peak varies with multiple parameters, but it can be confined in some limited regions and then located quickly by Lemma 2 and Corollary \ref{coro3}.

\begin{lemma}\label{lem3}
Assume $(\cos\theta_T,R_T)$ as the origin of the $\cos\theta-R$ plane, $\Delta f_m=-\Delta f_{M-1-m}$, $\Delta f_{0.5(M-1)}=0$ for odd $M$, then the FRB-resultant beampattern is symmetric with respect to the $R$-axis, conjugate symmetric with respect to the $\cos\theta$-axis, i.e.,
\begin{align}
B_{FRB}(\cos\theta,R)&=B_{FRB}(-\cos\theta,R)\nonumber\\
&=B_{FRB}^*(\cos\theta,-R).
\end{align}

\emph{Proof:} See Appendix F.$\hfill\blacksquare$
\end{lemma}

Basing on Lemma 3, we have an interesting corollary as follows.
\begin{corollary}
The amplitude of the FRB-resultant beampattern is symmetric with respect to the origin of the $\cos\theta-R$ plane if $\Delta f_m=-\Delta f_{M-1-m}$, i.e.,
\begin{align}\label{lem3cor}
|B_{FRB}(\cos\theta,R)|&=|B_{FRB}(-\cos\theta,R)|\nonumber\\
&=|B_{FRB}(-\cos\theta,-R)|\nonumber\\
&=|B_{FRB}(\cos\theta,-R)|.
\end{align}

\emph{Proof:} See Appendix G. $\hfill\blacksquare$
\end{corollary}
\begin{remark}
Lemma (\ref{lem3}) indicates that: when we optimize the sidelobe, we only need to focus on a quarter of the $\cos\theta-R$ plane, which significantly decreases the complexity of the following optimization problems.
\end{remark}


\section{PLS Performance Optimization}
In this section, we study the PLS-oriented FRB optimization problems by first formulating the achievable SR. We study two cases of interests, i.e., without/with the location information of eavesdropping, both turn out to be NP-hard. We first study the case without eavesdropping location information, and propose a SOA to minimize the sidelobe peak of the FRB-resultant beampattern outside the DARR. We then study the case with eavesdropping location information, and propose a BCDLA to find the optimized FOI vector.

\subsection{Problem Formulation}
The proposed FRB has a deterministic beampattern for a given FOI vector $\Delta \mathbf{f}$. Since Eve may be located at any region except DARR, i.e., target region $(\theta_T,R_T)$, the beampattern at Eve's location is not constant. When Eve locates exactly in the maximum sidelobe peak of the FRB-resultant beampattern, the situation of information leakage is the worst. Therefore, we need to design a FRB strategy to choose the FOI vector appropriately for better SR performance.

In order to study the performance of FRB, we temporally ignore the effects of transmission power and path-loss factor. The signal-to-noise ratio (SNR) at any region is defined as
\begin{align}
SNR(\theta,R)=\frac{|B_{FRB}(\theta,R)|^2}{\sigma^2}.
\end{align}
The achievable rates of Bob and Eve are expressed as
\begin{align}
R_{Bob}=\log\Big(1+\frac{|B_{FRB}(\theta_T,R_T)|^2}{\sigma_{T}^2}\Big),
\end{align}
\begin{align}
R_{Eve}=\log\Big(1+\frac{|B_{FRB}(\theta,R)|^2}{\sigma_{E}^2}\Big),
\end{align}
where $\sigma_{T}^2$ and $\sigma_{E}^2$ represent the powers of AWGN at Bob and Eve, respectively.

The SR is defined as the difference of rates between the legitimate channel and the wiretap channel, i.e.,
\begin{align}
R_S=
\begin{cases}
R_{Bob}-R_{Eve}&\text{if}\ R_{Bob}>R_{Eve}\\
0&\text{otherwise}
\end{cases}.
\end{align}
Then the SRM problem is formulated as follows,
\begin{align}
&\max_{\substack{\Delta \mathbf{f}}} R_S=R_{Bob}-R_{Eve}\nonumber\\
&s.t. \quad|\Delta f_m|\leq\Delta f_{max},\forall m\tag{P1}.
\end{align}

In the proposed FRB transmission scheme, the level of resultant beampattern is mainly determined by the FOI vector $\Delta \mathbf{f}$ and the array parameters. We assume that Bob can be located accurately and the signal energy of the confidential information is intentionally focused on this region. Thus, the resultant beampattern gain of Bob is constant and maximum. As the location of Eve is unknown, the worst SR will happen when Eve locates in the maximum sidelobe peak. Thus, the SRM depends on the peak level of sidelobe outside the DARR. To this end, the SRM problem (P1) boils down to find the optimal $\Delta\mathbf{f}$ to minimize the sidelobe level, which is expressed as follows,
\begin{align}
&\min_{\substack{\Delta \mathbf{f}}}  |B_{FRB}(\theta,R)|^2 \nonumber\\
&s.t. \quad|\Delta f_m|\leq\Delta f_{max},\forall m\nonumber\\
&\quad\quad\, (\theta,R)\notin\Omega_{m}\tag{P2},
\end{align}
where
\begin{align}
&|B_{FRB}(\theta,R)|^2=B_{FRB}^*(\theta,R)B_{FRB}(\theta,R)\nonumber\\
&=\frac{\sum\limits_{m=0}^{M-1}\sum\limits_{n=0}^{M-1}e^{j2Na(n-m)}\frac{\sin(Na-Nb\Delta f_m)\sin(Na-Nb\Delta f_n)}{N^2\sin(a-b\Delta f_m)\sin(a-b\Delta f_n)}}{M^2}\nonumber\\
&=\frac{\sum\limits_{m=0}^{M-1}\sum\limits_{n=0}^{M-1}\frac{\cos(2Na(n-m))\sin(Na-Nb\Delta f_m)\sin(Na-Nb\Delta f_n)}{N^2\sin(a-b\Delta f_m)\sin(a-b\Delta f_n)}}{M^2}\nonumber\\
&=\frac{\sum\limits_{m=0}^{M-1}\sum\limits_{n=0}^{M-1}\cos(2Na(n-m))g(\Delta f_m)g(\Delta f_n)}{M^2},
\end{align}
where $g(\Delta f_m)$ is the beampattern of the $m$-th FDSA with $N$ elements and the FOI $\Delta f_m$, expressed as follows,
\begin{align}
g(\Delta f_m)=\frac{\sin(Na-Nb\Delta f_m)}{N\sin(a-b\Delta f_m)}\label{E1}.
\end{align}

Therefore, the problem (P2) can be equivalently expressed as
\begin{align}
&\min_{\substack{\Delta 
\mathbf{f}}}\sum\limits_{m=0}^{M-1}\sum\limits_{n=0}^{M-1}\cos(2Na(n-m))g(\Delta f_m)g(\Delta f_n)\nonumber\\
&s.t. \quad|\Delta f_m|\leq\Delta f_{max},\forall m\tag{P3}.
\end{align}

Our target is to achieve optimal SR for Bob with the proposed FRB under two scenarios of interests, namely without/with location information of eavesdropping. In subsequent subsections, we propose two optimization algorithms for these scenarios.

\subsection{FRB-SR Optimization without Eavesdropping Location Information }
In practice, the location of eavesdropping is generally unknown. In order to implement FRB against unlocalized eavesdropping, the sidelobe of FRB-resultant beampattern should be as low as possible. In such case, the objective of SRM problem (P1) is to minimize the maximum sidelobe peak, and the optimization problem is
\begin{align}
&\min_{\substack{\Delta \mathbf{f}}} \max_{\substack{\theta,R}}|B_{FRB}(\theta,R)|^2 \nonumber\\
&s.t. \quad|\Delta f_m|\leq\Delta f_{max},\forall m\nonumber\\
&\quad\quad\  (\theta,R)\in\Omega_{ms}\tag{P4}.
\end{align}

Since the beampattern gain varies with the location ($\theta,R$) and FOI vector $\Delta \mathbf{f}$, (P4) is NP-hard and difficult to solve. To this end, an efficient algorithm named as SOA will be introduced to solve the problem.

\begin{algorithm}
\caption{Seeker Optimization Algorithm (SOA)}
\begin{algorithmic}[1]
\REQUIRE $M$, $N$, $f_c$, $d$, $\Delta f_{max}$, $\theta_T$, $R_T$
\ENSURE Optimized FOI vector $\Delta \mathbf{f}$
\STATE Initialize: position $\Delta\mathbf{F}$, search direction $\mathbf{D}$, step length $\mathbf{V}$, best personal position  $\mathbf{P}$, best global position $\mathbf{g}$
\FOR{$t=1$ to $T$}
\FOR{$s=1$ to $S$}
\STATE $W=0.9-0.8t/T$
\STATE Determine EG direction $\mathbf{D}_s$ by (\ref{EquEG})
\STATE Determine step length $\mathbf{V}_s$ by (\ref{EquLEN})
\STATE $\Delta\mathbf{F}_s^t=\Delta \mathbf{F}_s^{t-1}+\mathbf{D}_s\diamond\mathbf{V}_s$
\IF{$|\Delta\mathbf{F}_{s,j}^t|>\Delta f_{max}$}
\STATE $\Delta\mathbf{F}_{s,j}^t=\Delta f_{max}\times \mathrm{sgn}(\Delta\mathbf{F}_{s,j}^t)$
\ENDIF
\IF {Fitness($\Delta\mathbf{F}_s^t$)$<$Fitness($\mathbf{P}_{s}$)}
\STATE $\mathbf{P}_{s}=\Delta\mathbf{F}_s^t$
\ENDIF
\IF{Fitness($\Delta\mathbf{F}_s^t$)$<$Fitness($\mathbf{g}$)}
\STATE $\mathbf{g}=\Delta\mathbf{F}_s^t$
\ENDIF
\ENDFOR
\ENDFOR
\RETURN $\Delta \mathbf{f}=\mathbf{g}$
\end{algorithmic}
\end{algorithm}

\emph{1) Seeker Optimization Algorithm }

We adopt SOA to find an approximate optimization solution for problem (P4). SOA is a population-based heuristic search algorithm, which attempts to simulate the action of human's intelligent search for real-parameter optimization with their memory, experience and uncertainty reasoning \cite{Dai2006Seeker,4912428,5229209}.  The individual of this population is called seeker. An empirical gradient (EG) search direction and a step length are calculated to update seeker's position in each iteration. SOA shows better performance than existing algorithms, such as particle swarm optimization algorithm, genetic algorithm, in the optimization of interference suppression of linear antenna arrays \cite{Guney2011Seeker}.

As shown in Algorithm 1, SOA operates on a position matrix $\Delta\mathbf{F}\in\mathbb{Z}^{M\times S}$, which is treated as $S$ potential solutions or FOI vectors for problem (P4). Line 1 is used to initialize the seeker's position $\Delta\mathbf{F}$, search direction $\mathbf{D}\in\mathbb{Z}^{M\times S}$ and step length $\mathbf{V}\in\mathbb{Z}^{M\times S}$, best personal position $\mathbf{P}\in\mathbb{Z}^{{M\times S}}$ and best global position $\mathbf{g}\in\mathbb{Z}^{{M\times 1}}$. Line 4 is used to decrease the weight from 0.9 to 0.1 with the increasing of the iteration number so as to gradually improve the search precision. The search space of SOA is treated as a gradient field, and an EG can be determined to guide the seeker's search direction by evaluating his or his neighbor's current or historical position change \cite{4912428,5229209}. Line 5 determines the EG direction of the $s$-th seeker, which is expressed as
\begin{align}\label{EquEG}
\begin{cases}
&\mathbf{D}_s=\mathrm{sgn}(r_1\mathbf{d}_{ego}+r_2\mathbf{d}_{alt}+W\mathbf{d}_{pro})\\
&\mathbf{d}_{ego}=\mathrm{sgn}(\mathbf{P}_{s}-\Delta\mathbf{F}_s^t)\\
&\mathbf{d}_{alt}=\mathrm{sgn}(\mathbf{g}-\Delta\mathbf{F}_s^t)\\
&\mathbf{d}_{pro}=\mathrm{sgn}(\mathrm{Fitness}(\Delta\mathbf{F}_s^t)-\mathrm{Fitness}(\mathbf{P}_s))
\end{cases},
\end{align}
where $r_1$ and $r_2$ are random numbers uniformly distributed on the interval [0,1], sgn($\cdot$) is a signum function returning the value -1, 0 or 1. The subscript $s$ of matrix notation denotes the $s$-th column vector of matrix. The seekers use three EGs, i.e., proactive, egotistic and altruistic to determine their direction by evaluating their own current or historical positions or their neighbors. Proactive direction $\mathbf{d}_{pro}$ means the seekers may be proactive to change his search direction, and exhibits a goal-directed behavior by predicting and guiding the future behavior according to their past behaviors. Egotistic direction $\mathbf{d}_{ego}$ denotes the seekers go towards their historical best position through cognitive learning. Altruistic direction $\mathbf{d}_{alt}$ represents the seekers cooperate explicitly and go towards the desired goal by the social learning.

Line 6 of Algorithm 1 computes the search step length $\mathbf{V}$ of the $s$-th seeker, which is expressed as
\begin{align}\label{EquLEN}
\mathbf{V}_s=W|\mathbf{g}-\Delta\mathbf{F}_{rand}|\sqrt{-\ln(u+(1-u)r_3)}
\end{align}
where $\mathbf{g}$ and $\Delta\mathbf{F}_{rand}$ are the best seeker and a randomly selected seeker from $\Delta\mathbf{F}$, respectively, $r_3$ is a random number. $u$ is the input of fuzzy reasoning module, following the fuzzy rule ``If {$fitness\ value\ is\ small$}, then {$step\ length\ is\ short$}'', expressed as
\begin{align}
u=u_{max}-\frac{S-I_s}{S-1}(u_{max}-u_{min})
\end{align}
where $I_s$ is the number of $\Delta\mathbf{F}$ after sorting the fitness values. $u_{max}$ and $u_{min}$ denote the maximum membership degree value and minimum membership degree value, respectively. In this paper, we set $u_{max}=0.95$, $u_{min}=0.0111$ \cite{Dai2006Seeker}. Thanks to the fuzzy rule, the better seeker position is, the shorter the step length will be. Lines 7-10 update the seeker position $\Delta\mathbf{F}_s^t$. Lines 11-16 update the personal best position $\mathbf{p}_{s,b}$ and global best position $\mathbf{g}_b$ according to the new position $\Delta\mathbf{F}_s^t$. In Algorithm 1, the fitness function is the maximum sidelobe peak with $\Delta \mathbf{f}=\Delta\mathbf{F}_s^t$, $\theta\neq\theta_T$, $R\neq R_T$, expressed as
\begin{align}\label{FitFun}
&\mathrm{Fitness}(\Delta\mathbf{F}_s^t)=\max_{\substack{\theta,R}} |B_{FAB}(\theta,R)|^2 \nonumber\\
&s.t.\ \Delta \mathbf{f}=\Delta\mathbf{F}_s^t,\theta\neq\theta_T,R\neq R_T.
\end{align}

\emph{2) Complexity Analysis of SOA }

The computational complexity of Algorithm 1 is $\mathcal{O}(T\times S\times F)$, where $F$ is the complexity of fitness function in (\ref{FitFun}). According to Corollary \ref{coro3}, assuming $d=0.5\lambda$, the number of potential sidelobe peak is $2M\times N$ for even $M$ or $M\times N$ for odd $M$. Thus, the computational complexity of SOA becomes $\mathcal{O}(T\times S\times M\times N)$.

Since directly solving the (P4) is difficult when the eavesdropping location is unknown, we have to use SOA, a heuristic search algorithm, to find the optimal FOI vector. In next subsection we will study the SR optimization problem for FRB with the location information of eavesdropper, and a more efficient algorithm will be proposed to solve this problem.

\subsection{FRB-SR Optimization with Eavesdropping Location Information}

If we have known that the eavesdropper locates in a specific location $(\theta_E,R_E)$, the objective of (P3) is to find a set of suitable FOI to generate the minimum beampattern gain in the eavesdropping region, which is also NP-hard. Here, we resort to the block coordinate descent (BCD) method \cite{BSUM2013,Wen2012BCD,Tseng2001BCD} to handle the non-convex non-smooth FRB-SR problem with multi-block variables. Note that BCD has been employed to solve the complicated power allocation problem in wireless communication systems \cite{5756489}.

\emph{1) A Brief Introduction of BCD Method}

Consider the following optimization problem,
\begin{align}
&\min_{\substack{\Delta\textbf{f}}}\quad f(\Delta f_0,\ldots,\Delta f_{M-1})\nonumber\\
&\text{s.t.}\quad \Delta f_m\in \mathcal{X}_m,m=0,2,\ldots,M-1,
\end{align}
where $\Delta\textbf{f}=[\Delta f_1,\cdots,\Delta f_{M-1}]^T\in\mathbb{R}^{M\times1}$, $\mathcal{X}_m\subseteq\mathbb{R}$ is a closed convex set, and $f:\prod_{i=0}^{M-1}\mathcal{X}_m\rightarrow\mathbb{R}$ is a continuous function. The BCD method is one of the efficient approaches to solve the above optimization problem \cite{BSUM2013}. At each iteration of this method, the function is minimized only with respect to a single block of variables while the rest of the blocks are held fixed. More specifically, in the $r$-th iteration of the algorithm, $\Delta\textbf{f}$ is updated as follows
\begin{align}\label{BCDRULE}
\begin{cases}
\Delta f_m^r=\arg\min_{\substack{\Delta f_m\in\mathcal{X}_m}} f(\Delta f_m,\Delta\mathbf{f}_{-m}^{r-1})\\
\Delta f_n^r=\Delta f_n^{r-1},\forall n\neq m, m=r\ \mathrm{mod}\ M.
\end{cases}
\end{align}
where $\Delta\mathbf{f}_{-m}^{r-1}\overset{\Delta}{=}[\Delta f_0^{r-1},\cdots,\Delta f_{m-1}^{r-1},\Delta f_{m+1}^{r-1},\cdots,\Delta f_{M-1}^{r-1}]^T\in\mathbb{R}^{(M-1)\times1}$ is the remaining variables in $\Delta\mathbf{f}^{r-1}$ after removing $\Delta f_m^{r-1}$, and $f(\Delta f_m,\Delta\mathbf{f}_{-m}^{r-1})=f(\Delta f_1^{r-1},\cdots,\Delta f_{m-1}^{r-1},\Delta f_m,\Delta f_{m+1}^{r-1},\cdots,\Delta f_{M-1}^{r_1})$ \cite{Tseng2001BCD}.

\emph{2) BCD Algorithm of (P3) }

Following the updating rule of BCD in (\ref{BCDRULE}), we assume that the FOI $\Delta f_m$ is updated in the $r$-th iteration by solving the following problem
\begin{align}
&\min_{\substack{\Delta f_m}}g^2(\Delta f_m)+\sum\limits_{\substack{n=0\\n\neq m}}^{M-1}\cos(2Na(n-m))g(\Delta f_n^{r-1})g(\Delta f_m)\nonumber\\
&\text{s.t.}\quad |\Delta f_n|\leq \Delta f_{max},\forall n.\tag{P5}
\end{align}

Let us define the function $g(\Delta f_m)$ as a single block variable. Obviously, (P5) is a typical convex optimization problem of quadratic function with one variable. It is easy to find that the optimal solution of problem (P5) is expressed as
\begin{align}
g(\Delta f_m)=\Bigg[-\sum\limits_{\substack{n=0\\n\neq m}}^{M-1}\frac{\cos(2Na(n-m))g(\Delta f_n^{r-1})}{2}\Bigg]_{g_{min}}^{g_{max}}\label{E2}
\end{align}
where $[\cdot]_{g_{min}}^{g_{max}}$ denotes the projection onto $[g_{min},g_{max}]$. $g_{min}$ and $g_{max}$ are the minimum and the maximum value of the function $g(\Delta f_m)$, respectively.

\begin{algorithm}
\caption{Block Coordinate Descent Linear Approximation (BCDLA)}
\begin{algorithmic}[1]
\REQUIRE Array Parameters $N$, $M$, $f_c$, $d$, $\theta_T$, $R_T$ \\ \qquad Eve's location $(\theta,R)$
\ENSURE Optimized frequency offset $\Delta \mathbf{f}$
\STATE Initialize: $\Delta \mathbf{f}^0$, $r=0$, $\varepsilon$, $r_{stop}$
\REPEAT
\STATE $m=r\,\mathrm{mod}\,M$, $r=r+1$
\STATE Compute $g(\Delta f_m)$ by equation (\ref{E2})
\STATE Select suitable peak points $P_{neg}$ and $P_{pos}$
\STATE Linear approximation: $g(\Delta f)\approx A_0\Delta f+A_1$
\STATE $\Delta f_m=\frac{g(\Delta f_m)-A_1}{A_0}$
\STATE $\Delta f_n^r=\Delta f_n^{r-1},\forall n\neq m$
\UNTIL $|\Delta \mathbf{f}^{r}-\mathbf{f}^{r-1}|<\varepsilon$ or $r=r_{stop}$
\end{algorithmic}
\end{algorithm}

\emph{3) Linear approximation (LA) method}

\begin{figure}[!t]
\centering
\includegraphics[width=3.5in]{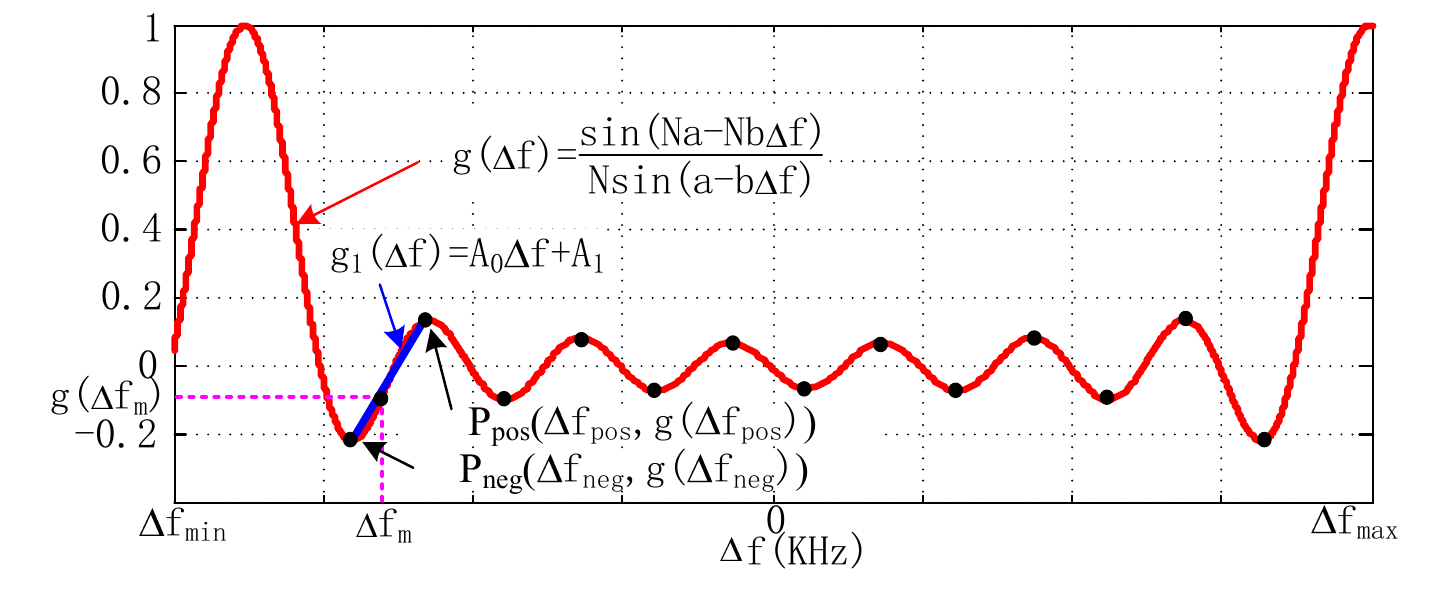}
\caption{Linear approximation method by a linear function with $f_c=60$GHz, $N=15$, $d=0.5\lambda$, $R_T=1000$m, $R=200$m, $\theta_T=90^0$, $\theta=20^0$.}
\label{t6}
\end{figure}
Utilizing (\ref{E1}) and (\ref{E2}), we can compute the FOI $\Delta f_m$  by decomposing $N$-fold angle formula of sine function and solving $N$-degree univariate polynomial equation. However, it is complex and difficult to solve when $N>4$. We use LA method to solve the solution $\Delta f_m$ of (\ref{E2}). As shown in Fig. \ref{t6}, we compute all the positive peak points and negative peak points from $\Delta f_{min}$ to $\Delta f_{max}$. Then, A suitable peak interval $[\Delta f_{neg},\Delta f_{pos}]$ is chosen according to the FOI constraint. A linear function $g_1(\Delta f)$ is used to approximate the function $g(\Delta f)$ in such interval, which is expressed as
\begin{align}
g(\Delta f)\approx g_1(\Delta f)=A_0\Delta f+A_1,
\end{align}
where $A_0=\frac{g(\Delta f_{pos})-g(\Delta f_{min})}{\Delta f_{pos}-\Delta f_{neg}}$, $A_1=g(\Delta f_{neg})-A_0\Delta f_{neg}$. Thus, we get the solution of (\ref{E2}) as follows
\begin{align}
\Delta f_m=\frac{g(\Delta f_m)-A_1}{A_0}\label{E4}
\end{align}

\emph{4) BCDLA Algorithm}

The BCDLA algorithm is summarized as the whole procedure to solve the problem (P5) as shown in Algorithm 2. Line 1 initializes the parameters, where $\varepsilon>0$ and $r_{stop}\in\mathbb{N}$ are the condition value of stopping BCDLA iteration. Line 4 computes the minimum $\Delta f_m$ with (\ref{E2}). Lines 5-6 are used to LA. Lines 7-8 update the variables. Line 9 is the stopping criterion of BCDLA.

\emph{5) Complexity Analysis of BCDLA}

The computational complexity of Algorithm 2 is $\mathcal{O}(r_{max})$, where $r_{max}$ is the maximum iteration. Compared with SOA, BCDLA is much simpler and efficient to find an optimal frequency offset for SRM on specific location.

\section{Simulation Results}
In this section, we will highlight the advantages of the proposed FRB schemes by comparing the SR performance with other reference schemes through numerical simulation.
\subsection{Simulation Assumption}
Unless otherwise specified, all simulations consider the linear array with isotropic antenna, the channel model follows Section \uppercase\expandafter{\romannumeral2}, the transmitter knows the location of Bob at $(\theta_T,R_T)=(90^0,500\mathrm{m})$. As shown in Fig. \ref{figPro}, in order to study the profile of SR performance, we assume Eve may move along three lines, i.e., the trajectory $T_{12}$ in angle-domain, the trajectory $T_{13}$ in range-domain and the trajectory $T_{14}$ in angle-range-domain, while keeping Bob's location as a constant. The system operates at 73 GHz with an average transmit power $P_T=40$ dBm. The FOI upper bound is $\Delta f_{max}\leq 10^{-5}f_c$. We also assume the noise power received by both Bob and Eve is $-100\mathrm{dBm}$, i.e., $\sigma_T^2=\sigma_E^2=10^{-10}\mathrm{mW}$.
\begin{figure}[!t]
\centering
\includegraphics[width=3.5in]{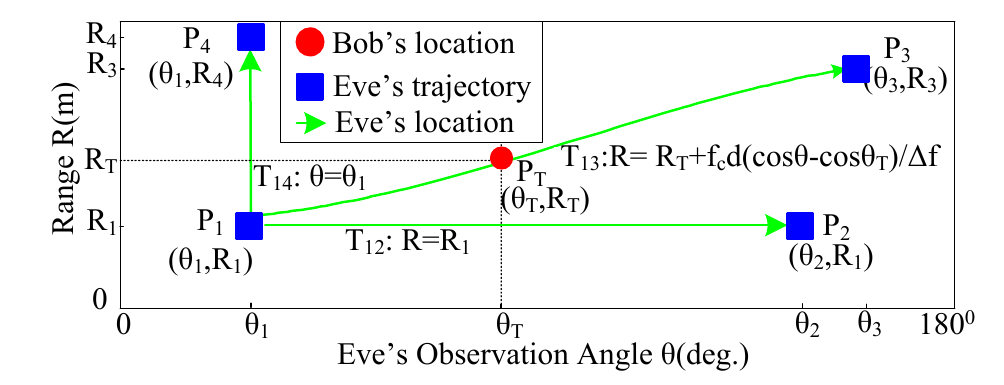}
\caption{Typical moving trajectory of eavesdropping for profile analysis of SR performance. Bob's location is fixed in ($\theta_T,R_T$). Eve has three types of moving trajectory, i.e., $T_{12}$ in angle-domain, $T_{14}$ in range-domain and $T_{13}$ in angle-range-domain. $\Delta f$ is the FOI of FDA-based RAB. }
\label{figPro}
\end{figure}
According to the NYC experimental result in  mmWave LoS communications \cite{6515173,6834753}, we adopt a log-distance model to model the path loss as
\begin{equation}
P_L(R)=\alpha+10n\log_{10}(R)[\text{dB}],
\end{equation}
where $\alpha=69.8$, $n=2$, $R$ denotes propagation distance in meter.

\subsection{FRB without Eavesdropping Location }
\begin{figure}[!t]
\centering
\includegraphics[width=3.5in]{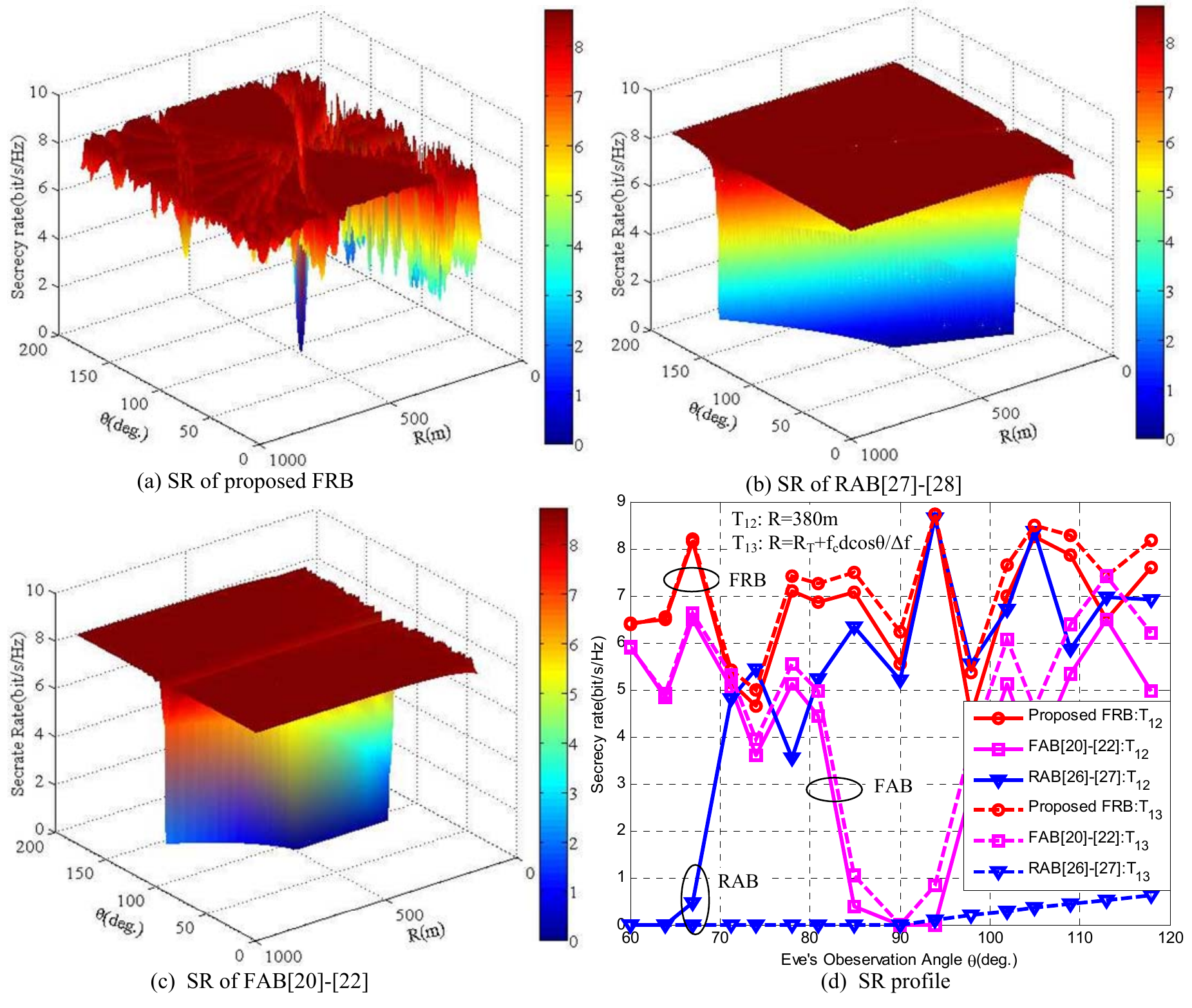}
\caption{SR comparison for different beamforming schemes without eavesdropping location with $N=11$, $M=15$. (a) SR of proposed FRB. (b) SR of RAB. (c) SR of FAB. (d) SR profile of proposed FRB, RAB and FAB on trajectory $T_{12}:R=380$m and $T_{13}:R=R_T+f_cd\cos\theta/\Delta f$. }
\label{t8}
\end{figure}
We use SOA algorithm to select the FOI vector for a minimum FRB-resultant sidelobe peak when the eavesdropper's location is unknown. As shown in Fig. \ref{t8}, the achievable SR of the proposed FRB is compared with existing FRB and RAB. Fig. \ref{t8}(a) shows that the SR of the proposed FRB generally keeps very high level except in the target region where Eve co-locates with Bob. Fig. \ref{t8}(b)-(c) demonstrate that the mainlobe paths of the existing FAB and RAB bring a long slit band, where the SR is almost zero and the information leakage is worst. Fig. \ref{t8}(d) is the SR profile near target angle in angle-domain. Assume Eve is along the trajectory $T_{12}$ or $T_{13}$ of Fig. \ref{figPro}, where $R=380$m or $R=R_T+f_cd\cos\theta/\Delta f$. The SRs of the proposed FRB along these two trajectories are higher and more robust than FAB and RAB. In addition, both FAB and RAB exist zero SR region along the mainlobe path. Outside the DARR, Because the beampattern of the proposed FRB is restrained by destructive superposition of subbeam, the mainlobe path that is inherent in FAB and RAB is not exist, thus the eavesdropper can not effectively intercept the confidential signal. Therefore, the proposed FRB eliminates the security threat of zero SR and keeps high SR outside the DARR. The SR performance of the proposed FRB is superior to the existing angular beamforming schemes.

\subsection{FRB with Eavesdropping Location}
\begin{figure}[!t]
\centering
\includegraphics[width=3.5in]{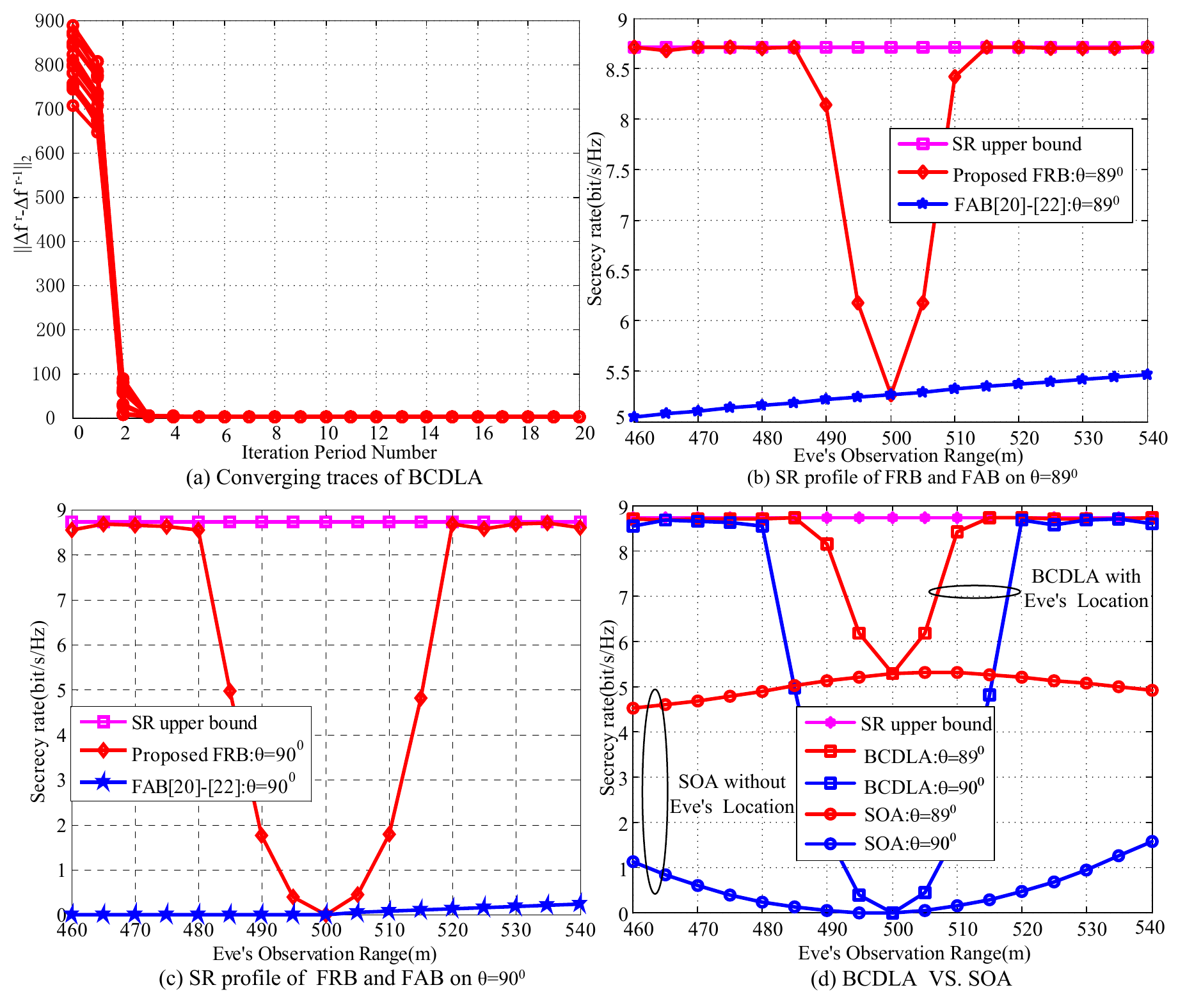}
\caption{Typical performance of BCDLA with $N=9$, $M=11$. (a) Converging traces at $\theta=30^0,\ R=200$m. (b) SR profile of FRB and FAB on $\theta=89^0$. (c) SR profile of FRB and FAB on $\theta=90^0$. (d) Optimization performance comparison with BCDLA and SOA. }
\label{t7}
\end{figure}
In order to keep the FRB-resultant sidelobe beampattern as low as possible at the eavesdropping location, we adopt the BCDLA algorithm to compute the FOI vector quickly for the proposed FRB. Fig. \ref{t7}(a) shows the converging process of the BCDLA algorithm, where each curve starts from a set of randomly selected FOIs. After about 4 iteration, the difference of FOI vectors between the $r$-th iteration and the $(r-1)$-th iteration converges to zero. Fig. \ref{t7}(b)-(c) compare the SRs of the proposed FRB and FAB with eavesdropping location, where eavesdropper moves along the trajectory $T_{14}$ of Fig. \ref{figPro} with $\theta=89^0$ or $90^0$. As FRB is angle-range-dependent while FAB is angle-dependent and range-independent, FRB can use eavesdropping location information to generate minimum beampattern gain on Eve's location. Thus, The SR of FRB is close to upper bound and superior to FAB in Fig. \ref{t7}(b)-(c). Fig. \ref{t7}(d) is optimization performance comparison between BCDLA and SOA. As expected, the SRs provided by BCDLA are superior to SOA. For example, in location $(89^0,460\mathrm{m})$, the SR of BCDLA reaches to 8.7 bit/s/Hz, while the SR of SOA is 4.5 bit/s/Hz which is only 51.7\% of BCDLA. Thus, BCDLA can effectively address the problem with known eavesdropper location.

\subsection{Impact of Antenna Number}

\begin{figure}[!t]
\centering
\includegraphics[width=3.5in]{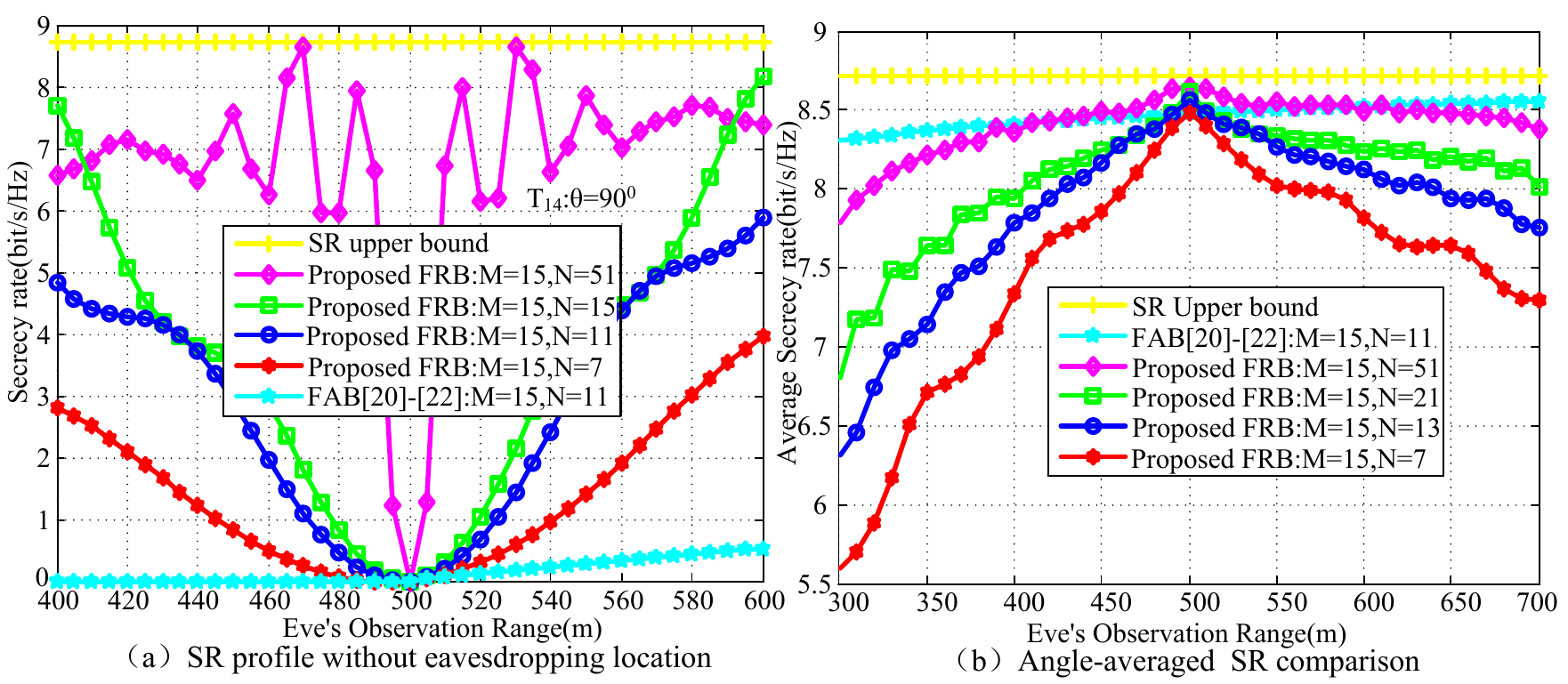}
\caption{SR comparison with different subarray antenna numbers when $M=15$, $R_T=500$m, $\theta_T=90^0$. (a) SR profile without eavesdropping location on trajectory $T_{14}:\theta=90^0$. (b) Angle-averaged SR comparison.}
\label{t9}
\end{figure}
We compare the SR performance with different subarray antenna number when $M=15$. As shown in Fig. \ref{t9}(a), assume Eve moves along trajectory  $T_{14}$ of Fig. \ref{figPro} with $\theta=90^0$, i.e., the target direction, the SR performance is improved as the subarray antenna number $N$ increases. This is because that the subarray sidelobe peak decreases when $N$ increases, which generates low sidelobe level of beampattern gain. Furthermore, the proposed FRB achieves higher SR than FAB, even the array number of FRB is less than FAB. Fig. \ref{t9}(b) demonstrates the angle-averaged SR\footnote{Angle-averaged SR denotes the average of SR in the angle domain when range is a constant, i.e., $\overline{R}_{S|R_E}=\sum\limits_{\theta=0}^{179}R_S(\theta,R_E)/180$.} on each observation range. As the number of subarray antenna $N$ increases, the angle-averaged SR of FRB also increases. Since the angle-averaged SR of FRB with $N=51$ is close to that of FAB with $N=11$, we think that outside the DARR, the proposed FRB can not only eliminate the zero SR threat, but also achieve good SR performance that is comparable to FAB.
\begin{figure}[!t]
\centering
\includegraphics[width=3.5in]{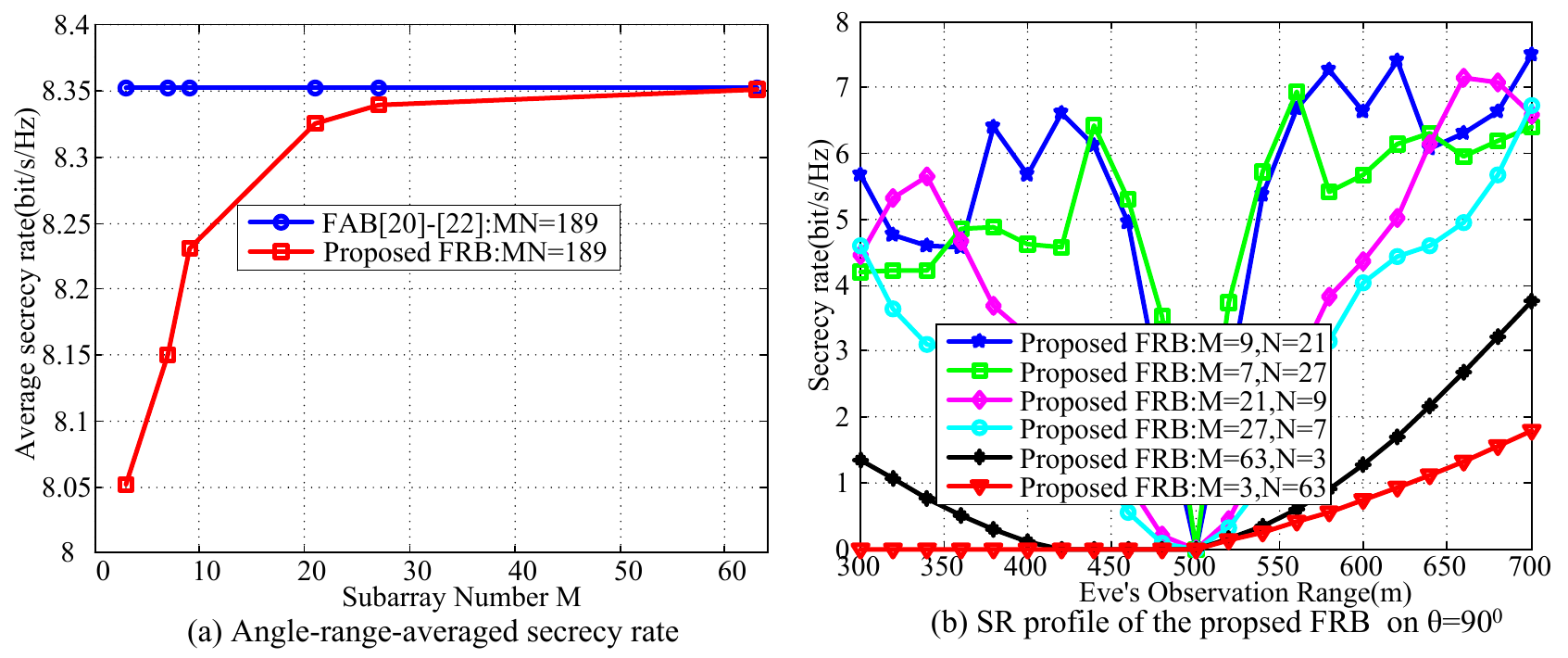}
\caption{SR comparison for different subarray combination with $MN=189$. (a) Angle-range-averaged SR. (b) SR profile of FRB on trajectory $T_{14}:\theta=90^0.$}
\label{t10}
\end{figure}

Fig. \ref{t10} makes a comparison of angle-range-averaged SR\footnote{Angle-range-averaged SR denotes the average of SR in the angle-range domain, i.e., $\overline{R}_S=\sum\limits_{\theta=0}^{179}\sum\limits_{R=1}^{2R_T}R_S(\theta,R)/(360\times R_T).$} within the same region for different subarray combination when $MN$ is a constant. As shown in Fig. \ref{t10}(a), the angle-range-averaged SR of FRB increases as $M$ increases. When $M$ increases to 27, the angle-range-averaged SR of FRB is approximate to FAB. However, a large subarray number $M$ is perhaps not optimal. As shown in Fig. \ref{t10}(b), in the target direction, the SR performance becomes better when the subarray number $M$ is close to the subarray antenna number $N$. According to (\ref{bfam}), large $M$ provides more DoF for the phase shift factor $\rho(\theta)$, which is beneficial to restrain the FRB-resultant sidelobe. On the contrary, the subbeam's mainlobe number and the subbeam's sidelobe level will increase as $M$ increases, which is disadvantageous to the the suppression of FRB-resultant sidelobe peak. Therefore, we should make a tradeoff between $M$ and $N$ when $MN$ is a constant.

\subsection{Impact of Maximum Frequency Offset Increment}
\begin{figure}[!t]
\centering
\includegraphics[width=3.5in]{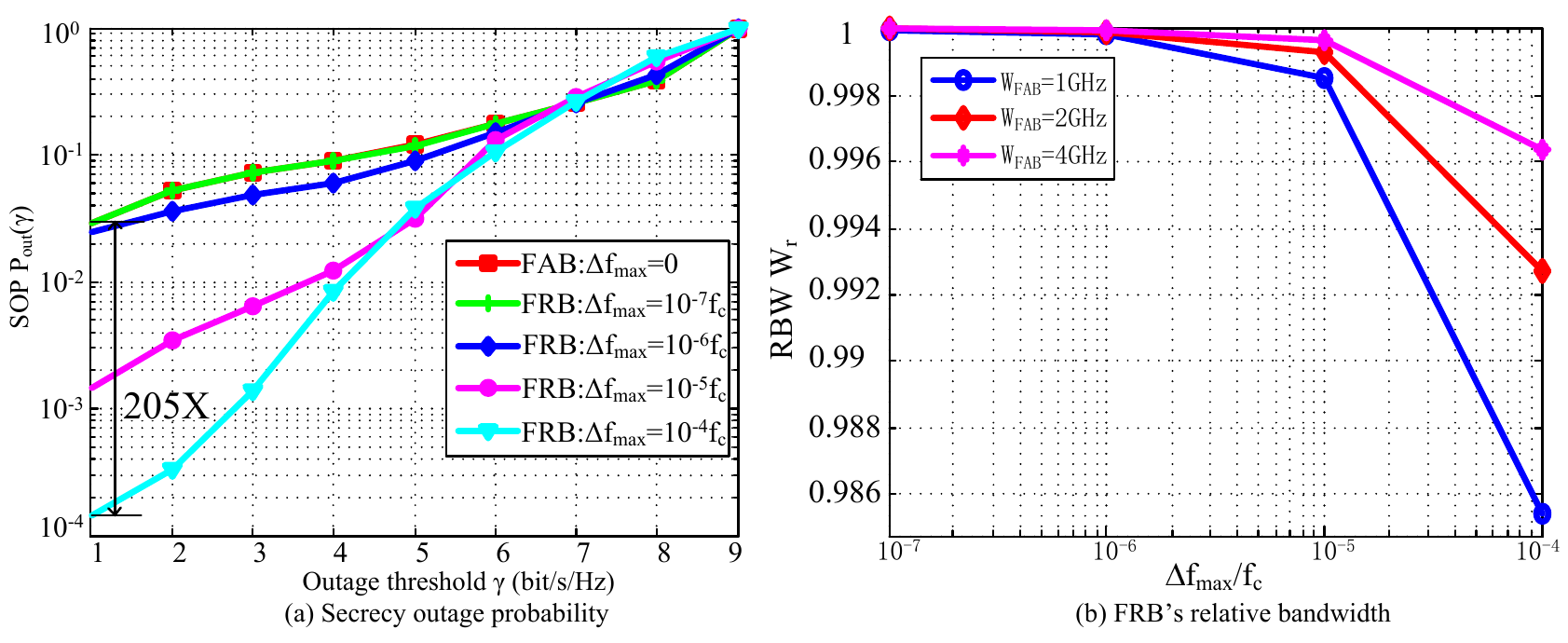}
\caption{Impact of $\Delta f_{max}$ with $0^0\leq\theta\leq179^0$, $50\mathrm{m}\leq R\leq2R_T$, $M=15$, $N=17$. (a) Secrecy outage probability. (b) FRB's relative bandwidth.}
\label{t11}
\end{figure}
Fig. \ref{t11} illustrates the impact of maximum FOI $\Delta f_{max}$ on the secrecy outage probability (SOP) and the bandwidth of the proposed FRB. Define the SOP as
\begin{align}
\mathrm{P}_{out}(\gamma)=\mathbb{P}(R_S<\gamma)
\end{align}
where $\gamma$ denotes the SR threshold.

Fig. \ref{t11}(a) shows that the SOPs of FRB decrease as the $\Delta f_{max}$ increases in low SR threshold. Especially, when FRB is simplified to FAB by setting $\Delta f_{max}=0$, the SOP of FAB is high to 205 times of FRB with $\Delta f_{max}=10^{-4}f_c$ when $\gamma=1$ bit/s/Hz. As $\gamma>6$ bit/s/Hz, the SOPs for different $\Delta f_{max}$ trend to unify and increase to 1. Since the proposed FRB uses the frequency offset to suppress the mainlobe path outside the DARR, high $\Delta f_{max}$ provides more DoF to eliminate the zero SR threat. Thus, the performance of SOP in low $\gamma$ improves with the increase of $\Delta f_{max}$.

Assume a double sideband system of FAB with bandwidth $W_{FAB}=2(f_h-f_c)$, where $f_h$ and $f_c$ are the upper bound frequency and carrier frequency, respectively. And the frequency spectrum range of FRB does not exceed the bandwidth of FAB. Then the relative bandwidth (RBW) of FRB is expressed as
\begin{align}
W_r=\frac{W_{FRB}}{W_{FAB}}=1-\frac{\Delta f_{max}}{f_h-f_c}
\end{align}

Fig. \ref{t11}(b) demonstrates that the RBW of the proposed FRB decreases with the increase of $\Delta f_{max}$. On the other hand, the decrease velocity of RBW will slow when the bandwidth of FAB increases. The RBW only decreases $0.15\%$ when $W_{FAB}=1$ GHz and $\Delta f_{max}=10^{-5}f_c$. In general, the order of magnitude of bandwidth in mmWave is at least gigahertz. Thus, the decrease of FRB bandwidth produced by frequency offset is very tiny and can be ignored in mmWave wireless communication systems.
\section{Conclusion}
In this paper, we propose a novel FRB scheme to improve PLS by exploiting the DoF of frequency offset supported by FDSA in mmWave band. Three transmission beamforming techniques, namely FAB, RAB and FRB, are studied within a unified FDSA architecture and instantiated by different setups of FOI. The SRM problems without/with eavesdropping location information are studied, where SOA and BCDLA algorithms are proposed to solve these complicated problems, respectively. According to the theoretical analysis and simulation results, we show that the proposed FRB achieves better SR performance than existing angular beamforming methods, i.e., FAB and RAB. The regime of low SR is eliminated completely in our proposed scheme when $\Delta f_{max}\leq10^{-4}f_c$, which solves the problem of zero SR in the mainlobe direction that is inherent in most of the existing beamforming schemes. Thus the SOP of FRB is about 205 times of FAB in $R_s<1$ bit/s/Hz. The FRB-SR performance is improved by increasing of the subarray number $M$, subarray element $N$ and maximum FOI $\Delta f_{max}$. In the future, we will further study the AN generated by the proposed FRB for higher PLS performance, and the multi-user secure transmission in some general scenarios.


%

\appendices
\section{Proof of Lemma 1}

\emph{Proof:} Rewrite the normalized RAB beampattern as
\begin{align}
B_{RAB}(\theta,R)&=\frac{\sin\left(\frac{L\pi f_cd(\cos\theta-\cos\theta_T)}{c}-\frac{L\pi\Delta f(R-R_T)}{c}\right)}{L\sin\left(\frac{\pi f_cd(\cos\theta-\cos\theta_T)}{c}-\frac{\pi\Delta f(R-R_T)}{c}\right)}\nonumber\\
&=\frac{\sin(Lx)}{L\sin(x)}
\end{align}
where $x=\frac{\pi f_c d(\cos\theta-\cos\theta_T)}{c}-\frac{\pi\Delta f(R-R_T)}{c}$.

Obviously, $B_{RAB}(\theta,R)$ is maximized when $\sin x=0$ or $x=2k\pi$($k=0$, $\pm1$, \ldots), which can be formulated as
\begin{align}
\text{max}\quad &B_{RAB}(\theta,R)\nonumber\\
s.t.\quad& x=2k\pi,k=0,\pm1,\cdots\label{APEQ1}.
\end{align}

From (\ref{APEQ1}), we can get $\sin(Lx)=\sin(2kL\pi)=0$ for integer $L\in[1,2,\cdots]$. According to the L'Hospital's rule, we have the maximum of $B_{RAB}(R,\theta)$ as follows,
\begin{align}
\text{max}\quad &B_{RAB}(\theta,R)=\lim\limits_{x\rightarrow 2k\pi}\frac{\sin(Lx)}{L\sin(x)}=1\nonumber\\
s.t.\quad& k=0,\pm1,\cdots,L=1,2,\cdots.
\end{align}

Therefore, the mainlobe gain value of RAB is 1 with suitable parameters $R$ and $\theta$. In the mainlobe path of RAB, we get following equation
\begin{equation}
x=\frac{\pi f_c d(\cos\theta-\cos\theta_T)}{c}-\frac{\pi\Delta f(R-R_T)}{c}=2k\pi\label{APEQ2}.
\end{equation}
Since $d\leq 0.5\lambda$ and $|\cos\theta-\cos\theta_T|\leq 2$, we can get
\begin{equation}
\left|\frac{f_cd(\cos\theta-\cos\theta_T)}{c}\right|\leq1\label{APEQ3}.
\end{equation}
Recall that we have assumed the conditions of mmWave LoS and $\Delta f\ll f_c$, then, $|\Delta f(R-R_T)|<c$, i.e.,
\begin{equation}
\left|\frac{\Delta f(R-R_T)}{c}\right|<1\label{APEQ4}.
\end{equation}
From (\ref{APEQ3}) and (\ref{APEQ4}), we have
\begin{equation}
\pi\left|\frac{f_cd(\cos\theta-\cos\theta_T)}{c}-\frac{\Delta f(R-R_T)}{c}\right|<2\pi\label{APEQ5}.
\end{equation}
According to (\ref{APEQ2}) and (\ref{APEQ5}), we have $k=0$. Then, we get following equation
\begin{equation}
\frac{\pi f_c d(\cos\theta-\cos\theta_T)}{c}-\frac{\pi\Delta f(R-R_T)}{c}=0\label{APEQ6},
\end{equation}
which can be simplified as
\begin{equation}
f_cd(\cos\theta-\cos\theta_T)=\Delta f(R-R_T)\label{APEQ7}.
\end{equation}

In $\cos\theta-R$ plane, we set the point $(\cos\theta_T,R_T)$ as the original point. The relation of propagation range $R$ and steering angle $\cos\theta$ is expressed as
\begin{equation}
R=\frac{f_cd}{\Delta f}\cos\theta\label{APEQ8}.
\end{equation}

Therefore, the angle $\varphi$ is expressed as
\begin{equation}
\varphi=\arctan\frac{f_cd}{\Delta f}.
\end{equation}

The proof of Lemma 1 is finished.$\hfill\blacksquare$

\section{Proof of Corollary 1}
\emph{Proof:} Noting that $\Delta f=0$, the rotated angle $\varphi=\frac{\pi}{2}$, i.e., the mainlobe of transmit beamforming is parallel with $R$-axis of $\cos\theta-R$ plane. As shown in Fig. \ref{fig4}, the beampattern of transmit beamforming is rotated to $\frac{\pi}{2}$ in $\cos\theta-R$ plane. Thus, the range parameter can't influence the transmit beamforming. The FDA-based RAB is degraded to CPA-based FAB.

The proof of Corollary 1 is finished.$\hfill\blacksquare$
\section{Proof of Corollary 2}
\emph{Proof:}Since $\Delta f_1=-\Delta f_2\neq 0$, thus, $\varphi_{1}=\arctan\frac{f_cd}{\Delta f_1}=\arctan\frac{f_cd}{-\Delta f_2}=-\varphi_{2}$. Therefore, the mainlobes of $\Delta f_1$ and $\Delta f_2$ will pass the target point $(\theta_T,R_T)$ and will be symmetric distribution along $R$-axis.

The proof of Corollary 2 is finished.$\hfill\blacksquare$
\section{Proof of lemma 2}
\emph{Proof:} In the region $\Omega_m$, all the mainlobes of subarray appear at the same position, i.e., $\theta=\theta_T$, $R=R_T$. Thus, the radiated fields of $M$ subarrays are superposed constructively, that is
\begin{equation}
B_{\Omega_m}^p=B_{\Omega_m}(\theta,R)|_{(\theta,R)=(\theta_T,R_T)}=1
\end{equation}

In the region $\Omega_{ms}$, there are $K$ mainlobe(s) and $M-K$ sidelobe(s). Therefore, the FRB-resultant beampattern can be expressed as
\begin{align}\label{Bomegams}
B_{\Omega_{ms}}(\theta,R)=&\frac{1}{MN}\bigg(\sum\limits_{k=1}^KNe^{j(M-1-2m_k)Na}+\sum\limits_{k=K+1}^{M}\nonumber\\
&\frac{e^{j(M-1-2m_k)Na}\sin(Na-Nb\Delta f_{m_k})}{\sin(a-b\Delta f_{m_k})}\bigg)
\end{align}
where $m_k\in\{0, 1, \cdots, M-1\}$. The first term on the RHS of (\ref{Bomegams}) is the summation of $K$ subarray mainlobes and the second term is the summation of the rest subarray sidelobes. When the phases of these subarrays in $\Omega_{ms}$ region are all coherent, i.e., $e^{j(M-1-2m_k)Na}$ is constant regardless the value of $m_k$, the beampatterns of FDSA can be effectively overlayed. Define the parameter $N^s$ as the maximum sidelobe peak of FDSA before normalization, thus $N^s\leq N$. Therefore, only when $M-K$ subarrays exhibit peak sidelobes and the phases of all subarrays are coherent, the FRB-resultant beampattern in $\Omega_{ms}$ region reaches the peak level $B_{\Omega_{ms}}^p$, which is expressed as
\begin{align}
B_{\Omega_{ms}}^p&=\frac{KN+(M-K)N^s}{MN}\nonumber\\
&\leq\frac{KN+(M-K)N}{MN}\nonumber\\
&=B_{\Omega_{m}}^p
\end{align}

In $\Omega_s$ region, there are only $M$ sidelobes. When these sidelobes are added coherently, the peak sidelobe level is written as
\begin{align}
B_{\Omega_s}^p&=\frac{MN^s}{MN}\leq\frac{KN+(M-K)N^s}{MN}=B_{\Omega_{ms}}^p
\end{align}

Thus, $B_{\Omega_{m}}^p\geq B_{\Omega_{ms}}^p\geq B_{\Omega_{s}}^p$.

The proof of Lemma 2 is finished.$\hfill\blacksquare$
\section{Proof of Corollary 3}
\emph{Proof:}According to (\ref{lem2}) of Lemma 2, the maximum sidelobe peak is confined in the $\Omega_{ms}$ region. Using (\ref{APEQ8}) in Appendix A, we have $R_m=\frac{f_cd}{\Delta f_m}\cos\theta$.

Rewrite the FRB-resultant beampattern as
\begin{align}
B_{FRB}(\theta,R)&=\mathbf{h}_{FRB}^*(\theta,R)\mathbf{w}_{FRB}(\theta_T,R_T)\nonumber\\
&=\sum\limits_{m=0}^{M-1}\frac{e^{j(M-1-2m)Na}\sin(Na-Nb\Delta f_m)}{MN\sin(a-b\Delta f_m)}\nonumber\\
&=\sum\limits_{m=0}^{M-1}e^{j\phi_m}A_m,
\end{align}
where $\phi_m=(M-1-2m)Na$, $A_m=\frac{\sin(Na-Nb\Delta f_m)}{MN\sin(a-b\Delta f_m)}$.
\begin{align}
&|B_{FRB}(\theta,R)|^2=(\sum\limits_{m=0}^{M-1}A_m\cos(\phi_m))^2+(\sum\limits_{m=0}^{M-1}A_m\sin(\phi_m))^2\nonumber\\
&=\sum\limits_{m=0}^{M-1}A_m^2+\sum\limits_{m=0}^{M-2}\sum\limits_{n=m+1}^{M-1}2A_mA_n\cos(\phi_m-\phi_n)\nonumber\\
&\leq\sum\limits_{m=0}^{M-1}A_m^2+\sum\limits_{m=0}^{M-2}\sum\limits_{n=m+1}^{M-1}2|A_mA_n|\nonumber\\
&=|\sum\limits_{m=0}^{M-1}A_m|^2
\end{align}

Thus, $e^{j\phi_m}=\pm1$ has more prominent effect for high sidelobe production than $e^{j\phi_m}\neq\pm1$. Therefore, we have $Na=n\pi$ for odd $M$ while $2Na=n\pi$ for even $M$. As $a=\frac{\pi f_cd\cos\theta}{c}$ when $(\theta_T,R_T)$ is set as the origin in $\cos\theta-R$ plane, we get $\cos\theta_n=\frac{\lambda n}{Nd}$ for odd $M$ or $\cos\theta_n=\frac{\lambda n}{2Nd}$ for even $M$.  The proof of Corollary \ref{coro3} is finished.
$\hfill\blacksquare$
\section{Proof of lemma 3}
\emph{Proof:} For simplicity, we assume $(\cos\theta_T,R_T)$ is the origin point, then, $\cos\theta_T=0$, $a=\frac{\pi f_cd\cos\theta}{c}$, $b=\frac{\pi R}{c}$. Thus, the FRB-resultant beampattern can be expressed as
\begin{align}
&B_{FRB}(a,b)=\sum\limits_{m=0}^{M-1}\frac{e^{j(M-1-2m)Na}\sin(Na-Nb\Delta f_m)}{MN\sin(a-b\Delta f_m)}\nonumber\\
&=B_0+\sum\limits_{m=0}^{\lfloor\frac{M-1}{2}\rfloor}\Bigg(\frac{e^{j(M-1-2m)Na}\sin(Na-Nb\Delta f_m)}{MN\sin(a-b\Delta f_m)}+\nonumber\\
&\quad\frac{e^{j(M-1-2(M-1-m))Na}\sin(Na-Nb\Delta f_{M-1-m})}{MN\sin(a-b\Delta f_{M-1-m})}\Bigg)\nonumber\\
&\overset{(i)}{=}B_0+\sum\limits_{m=0}^{\lfloor\frac{M-1}{2}\rfloor}\Bigg(\frac{e^{j(M-1-2m)Na}\sin(Na-Nb\Delta f_m)}{MN\sin(a-b\Delta f_m)}+\nonumber\\
&\quad\frac{e^{-j(M-1-2m)Na}\sin(Na+Nb\Delta f_m)}{MN\sin(a+b\Delta f_m)}\Bigg)\nonumber\\
&=B_{FAB}(-a,b)
\end{align}
where step (i) relies on the condition $\Delta f_m=-\Delta f_{M-1-m}$, $B_0$ is the beampattern of subarray $S_{0.5(M-1)}$ for odd $M$ defined as
\begin{equation}
B_0=
\begin{cases}
\frac{\sin(Na)}{MN\sin(a)}&\text{, odd}\ M\\
0&\text{, even}\ M
\end{cases}.
\end{equation}

Therefore, the FRB-resultant beampattern is symmetric with respect to the $R$-axis.
\begin{align}
&B_{FRB}^*(a,-b)\nonumber\\
&=B_0+\sum\limits_{m=0}^{\lfloor\frac{M-1}{2}\rfloor}\Bigg(\frac{e^{-j(M-1-2m)Na}\sin(Na-N(-b)\Delta f_m)}{MN\sin(a-(-b)\Delta f_m)}+\nonumber\\
&\quad\frac{e^{j(M-1-2m)Na}\sin(Na+N(-b)\Delta f_m)}{MN\sin(a+(-b)\Delta f_m)}\Bigg)\nonumber\\
&=B_{FRB}(a,b).
\end{align}

Thus, the FRB-resultant beampattern is conjugate symmetric with respect to the $\cos\theta$-axis.

The proof of Lemma 3 is finished. $\hfill\blacksquare$
\section{Proof of Corollary 4}
According to the Lemma \ref{lem3}, we get the following formula
\begin{align}
|B_{FRB}(\cos\theta,R)|&=|B_{FRB}(-\cos\theta,R)|\nonumber\\
&=|B_{FAB}^*(\cos\theta,-R)|.
\end{align}
Since $|B_{FRB}^*(\cos\theta,-R)|=|B_{FRB}(\cos\theta,-R)|$, thus we get equation (\ref{lem3cor}).

The proof of Corollary 4 is finished. $\hfill\blacksquare$



\ifCLASSOPTIONcaptionsoff
  \newpage
\fi

\bibliographystyle{IEEEtran}
\bibliography{IEEEabrv}

\end{document}